*Review*

# A Comprehensive Review of Myoelectric Prosthesis Control

**Mohammad Reza Mohebbian [1], Marjan Nosouhi [2], Farzaneh Fazilati [3], Zahra Nasr Esfahani [4], Golnaz Amiri [5], Negar Malekifar [6], Fatemeh Yusefi [7], Mohsen Rastegari [8] and Hamid Reza Marateb [9,10,*]**

[1] Department of Electrical and Computer Engineering, University of Saskatchewan, SK, Canada; mom158@usask.ca
[2] Biomedical Engineering Department, Engineering Faculty, University of Isfahan, Isfahan, Iran; mnosouhi@eng.ui.ac.ir
[3] Biomedical Engineering Department, Engineering Faculty, University of Isfahan, Isfahan, Iran; f.fazilati@eng.ui.ac.ir
[4] Biomedical Engineering Department, Engineering Faculty, University of Isfahan, Isfahan, Iran; zahranasr76@gmail.com
[5] Biomedical Engineering Department, Engineering Faculty, University of Isfahan, Isfahan, Iran; golnzamiri@gmail.com
[6] Biomedical Engineering Department, Engineering Faculty, University of Isfahan, Isfahan, Iran; negrmalekifar@gmail.com
[7] Biomedical Engineering Department, Engineering Faculty, University of Isfahan, Isfahan, Iran; ftmyou87@gmail.com
[8] Biomedical Engineering Department, Engineering Faculty, University of Isfahan, Isfahan, Iran; mohsen.rastegari@gmail.com
[9] Biomedical Engineering Department, Engineering Faculty, University of Isfahan, Isfahan, Iran; h.marateb@eng.ui.ac.ir
[10] Biomedical Engineering Research Centre (CREB), Automatic Control Department (ESAII) Universitat Politècnica de Catalunya-Barcelona Tech (UPC), Barcelona, Spain; hamid.reza.marateb@upc.edu

* Correspondence: h.marateb@eng.ui.ac.ir; Tel.: +98 (313) 793 5818.

**Abstract:** Prosthetic hands can be used to support upper-body amputees. Myoelectric prosthesis, one of the externally-powered active prosthesis categories, requires proper processing units in addition to recording electrodes and instrumentation amplifiers. In this paper, the following myoelectric prosthesis control methods were discussed in detail: On-off and finite-state, proportional, direct, and posture, simultaneous, classification and regression-based control, and deep learning methods. Myoelectric control performance indices, such as completion time and rate, throughput, lag, and path length, were reviewed. The advantages and disadvantages of the control methods were also discussed. Some of myoelectric prosthesis control's significant challenges are comfort, durability, cost, the application of under-sampled signals, and electrode shift. Moreover, the proposed algorithms must be usually tuned after each don and doff, which is not comfortable for the users. Real-time simultaneous and proportional myoelectric control, resampling human's arm, has brought much attention. However, increasing the degree of freedom reduces the overall performance. Applying a 3D printed prosthesis arm and under-sampled electromyographic signals could reduce the fabrication cost and improve the application of such methods in practice. There are many technological and clinical challenges in this area to reduce the prosthesis rejection rate.

## 1. Introduction

Amputation impacts patients both physically and mentally [1]. Not only does it involve the reduction of earned income throughout the lifetime, but it also affects the family, society, and health services [2]. Pathologies, such as catastrophic accidents, and traumatic malignancies, may cause amputations [3]. Depression can also contribute to the procedure following a spontaneous impairment [4,5].

Figures 1 and 2 indicate the YLD (years lived with disability) of unilateral and bilateral upper-limb amputations from 1990 to 2018 in various countries (Source: Institute for Health Metrics Evaluation. Used with permission. All rights reserved). The risk of limb loss increases with age [6]. Trauma is the leading cause of upper limb amputations, accounting for 80% of amputations in men aged 15 to 45. Cancer/tumors and vascular disease disorders are the second most common reasons [6].

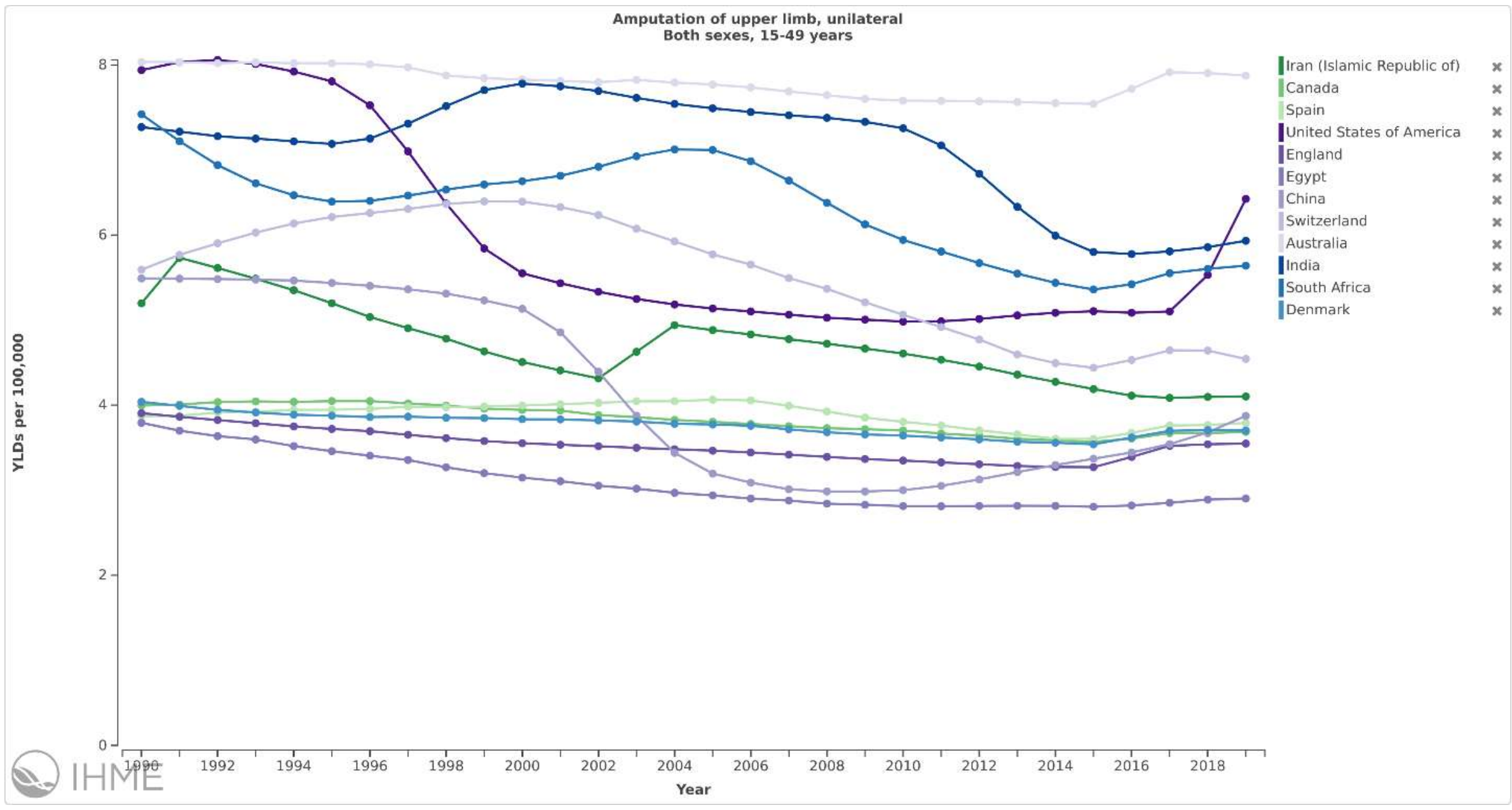

**Figure 1.** The YLD (years lived with disability) of unilateral upper-limb amputations for both sexes, 15-49 years of age per 100,000 population, based on the Global Burden of Diseases in various countries from 1990 to 2018 *[7]*.

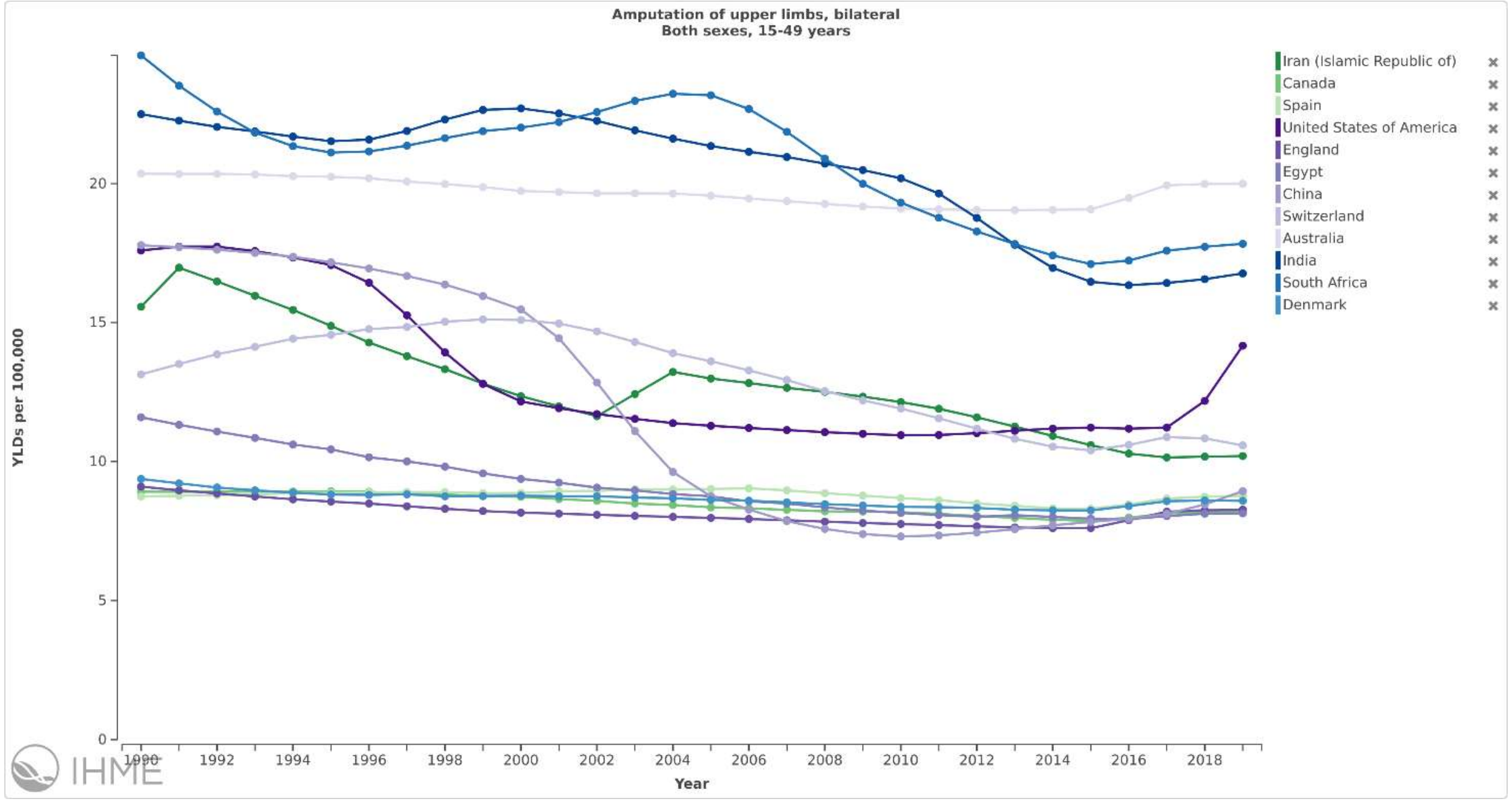

**Figure 2**. The YLD (years lived with disability) of bilateral upper-limb amputations for both sexes, 15-49 years of age per 100,000 population, based on the Global Burden of Diseases in various countries from 1990 to 2018 *[7]*.

Prosthetic hands can be used to support upper body amputees. The prosthetic hands could be categorized into the following: Body-powered, pneumatic-powered, and electric-powered, and some of those are complicated and costly. For example, an advanced electric-powered myoelectric hand with multiple gripping fingers can cost up to 50,000 USD which is not affordable due to lack of public health coverage [8–10]. The myoelectric prosthesis requires proper processing units besides surface electrodes and instrumentation amplifiers. There are various schemes for controlling myoelectric prosthesis. This paper reviews myoelectric-controlled upper-limb prostheses, focuses on their control technique, and designs low-cost prosthetics based on electromyographic surface (sEMG).

## 2. Materials and methods

### *2.1. State of the art for prosthetic design and fabrication*

The ability to do the whole spectrum of manual activities, from grasping to freely moving, is achieved by the upper limb's body organ. Consequently, amputation in this body organ would severely confine individuals in their lifetime. In general, the upper limb is divided into four different parts: shoulder, arm or brachium, forearm or antebrachium, and hand, all of which play a significant role in defining the level of amputation the consequent limitations in functions [11]. The TRANSCARPAL, wrist disarticulation (between the carpals and radius and ulna), TRANSRADIAL (through the long bones of the radius and ulna), elbow disarticulation (between the radius and ulna and humerus), TRANSHUMERAL (through the humerus), shoulder disarticulation (between the humerus and scapula), and forequarter (amputations closer to the center of the body) are the seven classes of this inauspicious phenomenon [12]. Figure 1 depicts parts of the upper limb accompanied by the classes of amputation and possible limitations.

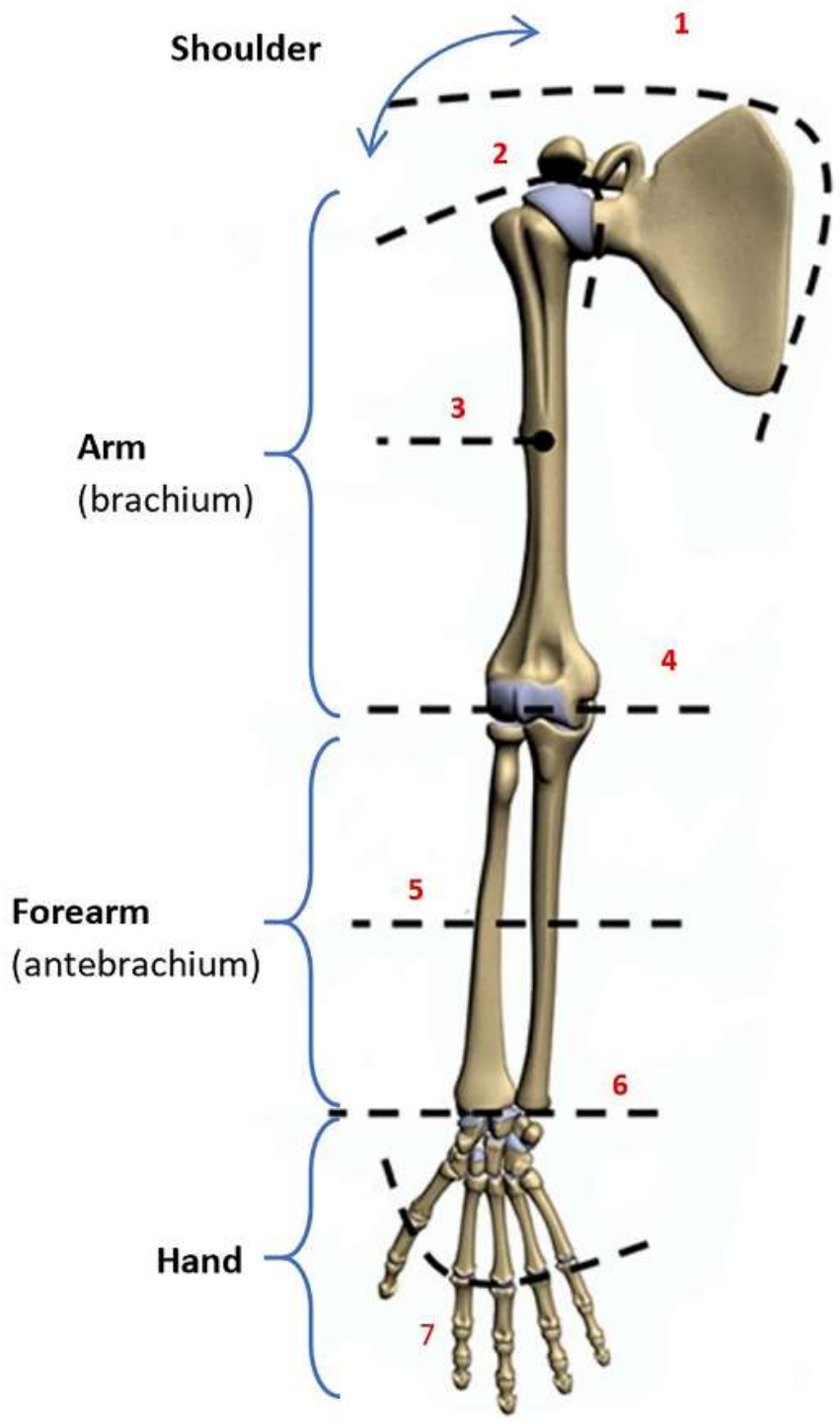

| NO | Upper limb amputation level | | Dysfunctions |
|---|---|---|---|
| 1 | Interscapular-thoracic (forequarter) | | •Loss of all arm/hand functions<br>•Partial or complete loss of clavicle and scapula |
| 2 | Shoulder disarticulation | | •Loss of all arm and hand function |
| 3 | TRANSHUMERAL | short | •Loss of internal and external rotation<br>•Loss of all arm and hand<br>•Partial or complete loss of clavicle scapular |
| | | long | •Loss of hand and wrist function<br>•All pronation and supination<br>•Most internal and external rotation<br>•Elbow flexion & extension |
| 4 | Elbow disarticulation | | Loss of:<br>•Hand and wrist function<br>•All pronation and supination<br>•Elbow flexion and extension |
| 5 | TRANSRADIAL | Short | Loss of:<br>•Hand and wrist function<br>•All pronation and supination |
| | | Long | •Loss of hand and wrist function<br>•Loss of most pronation and supination |
| 6 | Wrist disarticulation | | Loss of:<br>•Hand and wrist function<br>•About 50% pronation/supination |
| 7 | TRANSCARPAL | | •Loss of some or all grip function |

**Figure 3.** Level of upper limb amputation and the consequent dysfunction (Modified with permission from [12] and [6].

Considering the aesthetic issues and dysfunctions for the suffered individuals, the majority prefer to overcome the concerns by using terminal devices similar to a hand. The hand and the hook are two types of devices providing the required features. These devices are divided into two categories, namely the passive and the active hand. Although the former has no moving ability, they are considerably lighter, cheaper, and more long-lasting. The patient could benefit from some primary advantages of these products and solve the cosmetic issues while the fingers are fixed and without any advantages of grasping. The latter, however, could be controlled and bring back not all but some significant moving abilities and freedoms, such as limited grasping, movement of some fingers, and holding objects to the individuals. Both of the mentioned devices are produced in various features, ranging from size and materials to colors, notwithstanding these products' vulnerability to further damages [13].

The active devices themselves are divided into two categories, considering their features and abilities: first, the body-powered prosthesis in which the patient could control it with the limb's residual part. Not only are these devices comparatively light, but they can also be made waterproof. Moreover, the amputee can get a better sense of precision through the control cable's tension. However, these devices' drawbacks relate to the fact that the amputee has to have enough power and strength to prevail over the cable for the range of motion and set the device in a particular position. Second is the externally powered prosthesis, whose power is generated by the batteries within the toll. Such devices are controlled with a wide variety of inputs, such as electromyographic (EMG) signals, force-sensing resistors, and pull and push switches, among which the myoelectric prostheses are the most common [14]. These prostheses use the EMG signal from the contraction of the remaining muscles to operate movements. They can be controlled with much less trouble and much more freedom of movement. However, they are not as popular as they appear. The batteries and motors used in these devices are usually considered heavyweight devices, and they are not entirely waterproofed. Considering the power and maintenance, they should be recharged daily and used cautiously due to their delicate structures [15,16]. Figure 4 shows the summary of terminal devices.

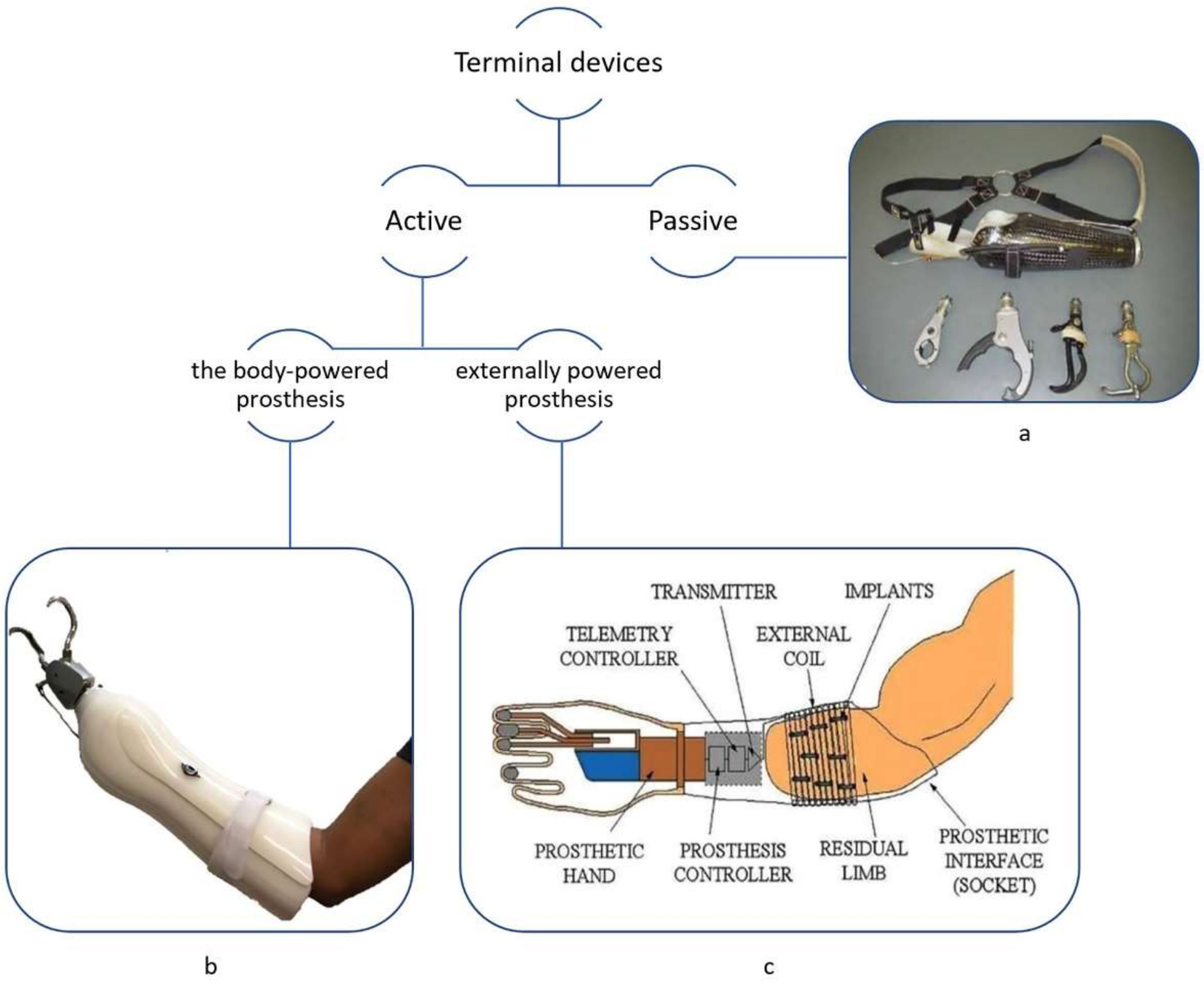

**Figure 4.** Terminal devices. Picture of (a) passive devices, (b) a body-powered prosthesis, and (c) the schematic of an externally powered prosthesis (Reproduced with permission from *[17–19]* ).

People usually have problems with these devices out of the attached electrode sensors to the limb's surface. The issues root from the fact that some of these limbs have fragile skins, or scar tissues, in which wearing these devices would deteriorate the situation by not fitting with the limb or sweating improperly. Furthermore, the change in the limb's size or shape and the lost connection among the electrodes and the limbs are reported as common difficulties. Apart from all, they are considerably expensive. Although many enhancements have been achieved to mitigate the problems, these disadvantages have made advanced prostheses unpopular and tend toward other products [15,16].

In order to address the existing problems with the usage of the myoelectric prosthesis, three novel methods have been introduced [17]:

- Implantable myoelectric sensor (IMES):
  Thanks to IMES's advent, it is possible to categorize and distinguish different recorded signals from small and large muscles. Consequently, both the accuracy and precision of the prosthesis will increase significantly [18].
- EMG pattern recognition (PR):
  EMG signals will be recorded from the remaining muscles or reinnervated muscles to identify the signals' patterns while the amputee is trying to do different tasks. Thanks to classifying these patterns, the accuracy of identification of the user’s motion, such as opening, griping, supination, pronation, and suchlike, increases considerably [19].
- Targeted Muscle Reinnervation (TMR):
  It is an invasive method in which the residual limb's remaining nerves are moved to the new muscles to be reinnervated. It can use these muscles and act as biological amplifiers for better controlling prosthetic devices [20].

Experts in the rehabilitation team tend to personalize each prosthesis concerning the amputee’s priorities and critical determinants such as physiology, power, and general status of the individual. Some would prefer switching techniques, among which the muscle co-contraction method, since more than one joint, could be involved. Considering the amputation level based on Figure 3, each person will contract the residual muscle(s) to control the device [17].

## 3. Results

The myoelectric prosthesis control can be seen from various points. The on-off control method is a simple technique that uses smoothed and rectified signals and a predefined threshold to open or close the prosthesis. In this control method, the prosthetic hand has a constant speed, and motors turn on and off and move at a fixed rate [21]. The term "proportional control" refers to a prosthetic control strategy where the mechanical output (force, velocity, position, or any function thereof) is continuously related to electromyographic (EMG) signals collected from remnant muscles of an amputee, which allows more precise and natural movements than is possible with on-off control [22]. Another field of view in prosthetic control is simultaneous control, which gives users real-time controllability, bringing smoothness and more satisfaction from prostheses [23]. Controlling prosthesis hand based on pattern recognition technique is one of the recent approaches designed based on feature engineering and classification and regression techniques [24]. Finally, advanced methods utilize Deep Learning approaches that do not require a feature selection step and are among the most accurate and robust approaches [25].

In the following section, we first review on-off and finite-state myoelectric control. Then, proportional, direct, and posture myoelectric control is investigated. Next, simultaneous myoelectric control methods are discussed. Pattern recognition-based methods, categorized into classification-based and regression-based, are then reviewed. Finally, myoelectric control using a deep learning approach is reviewed, and performance indices in myoelectric control are discussed.

*3.1. On-off and finite-state myoelectric control*

The finite state machine (FSM) is a method for detecting the purpose and coordinating hierarchical control of actions in neural prostheses, which is proper for a confined predefined number of movements. FSM includes four essential phases, analysis of input signals, estimating the succeeding state, starting the next state with refreshing the current state, and finally declaring the output control instruction [26].

The first reported use of myoelectric control was by Reiter [23]. In 1955, Battye *et al.* [27] demonstrated the feasibility of using myoelectric signals to control prostheses. They designed a system including electrodes, a suction pump for electrodes, a balanced amplifier, a discriminator that could apply power to the solenoid, close the split hook, and a power supply. The hook was controlled directly by the patient’s cerebral cortex. This system accomplished a simple open and close function when the user made a slight grasp. In further experiments, they decided to add some "backlash" into the switch's closing and opening so that the subject did not need to keep any extreme force.

In 1965, McKenzie [28] presented an artificial hand for patients with forearm amputation controlled by myoelectric potentials of the stump's forearm muscles. Muscle potentials of flexors and extensors muscles were the inputs of a two-channel low impedance amplifier. The output was an on-off control command for the opening and closing functions of the hand. The prosthesis movements were performed at a constant speed, and the grasp force could be modified just by timing. The artificial hand experimented on seven unilateral forearm amputations. The experiences showed that some subjects had difficulty isolating the contraction of forearm muscle groups, which resulted in the simultaneous signal entrance to both channels of amplifier that prevented sending the signal to motor switches and moving the hand.

In 1966, Dorcas and Scott [29] designed a three-state prosthesis control system using two levels of myoelectric activity at one site to enable on-off control of two separate actions. The first stage of the system, a differential input, is followed by a coupling circuit between two common-emitter steps and a common-collector step connected to a diode rectifier. The Rectifier’s output enters a Schmitt trigger circuit whose loads are two relays. Using electromechanical relays as switching components provided complete isolation between the control circuit and the controlled device. In this system, the user utilizes visual and acoustic feedback to specify the state of the prosthesis. An EMG signal with a small amplitude of about 250 μVp-p (sinusoidal signal) will activate the first relay. A larger amplitude of the input signal around 300 μVp-p will energize the second relay and deactivate the first one. Due to the time constants of the circuit, switching directly from a zero state to the situation in which just the second relay operates. Therefore, independent control of the relays is presented. They used an electric motor-driven hook to evaluate their control system. According to the clinical evaluation, the only remaining problem was the unintentional performance of the control system due to Simultaneous activity of the biceps and shoulder, which was expected to be solved by training.

Chappell *et al.* [30] developed a microprocessor-based controller for a single degree of freedom artificial hand manufactured by Viennatone of Austria. They used five surface electrodes to produce a single bipolar EMG signal. Two pairs of electrodes were used to record the EMG signal of flexor muscle and extensor muscle. The fifth electrode was the reference electrode. These two EMG signals were amplified with different gains and passed through a low-pass filter and a peak detector. According to input EMG signals and threshold values, the thumb and fingers could be closed, opened, or held in their position to grasp an object. Contracting the extensor muscles would open the hand, and contracting the flexor muscles resulted in its closure. If both extensor and flexor muscles generated signals that were lower than a fixed value, no voltage would apply to motors, and the prosthesis was static. Using feedback signals of position, touch, and slip sensors enhanced the control system and made the grasping function easier.

In 1995, Saxena *et al.* [31] developed and examined a two-channel functional electrical stimulation system to improve on-off control of grasping function in people with tetraplegia. The system was tested on subjects who had some remained wrist tension and paralyzed yet innervated finger flexors with almost normal shoulder and elbow coordination. The electromyographic control signal was recorded using surface electrodes located on the wrist extensors. After amplifying, rectifying, and integrating this signal, it was compared to a threshold for implementing finger and thumb flexors' on-off control. The test results showed that using this system enhanced the strength of the grasp in subjects, provided good performance reliability, and did not cause any side effects. However, functional tests showed that some subjects could not correctly use this system due to their physiological and anatomical problems.

In 2007, Sanches *et al.* [32] developed an EMG acquisition and processing system using a single field-programmable analog array (FPAA), which facilitated the recording of low amplitude biopotentials with high common-mode interferences. The reference EMG electrode and active electrodes are connected to the reference pin of FPAA and chopper amplifier inputs, respectively. The filter stage was achieved using a second-order Chebyshev band-pass filter and a bilinear filter. The bandwidth of this stage was limited from 50 to 350 Hz according to EMG signal characteristics. The calculation of the signal's RMS value was implemented using analog blocks of multiplication, square root derivation, and an external RC (which was used as a first-order filter to reduce the RMS signal ripple) of the FPAA. The last stage of this system, a comparator circuit, generated a trigger pulse (on/off control), which could be used as a command for external devices. The application of interference minimization methods provided a clean signal with a sufficient signal-to-noise ratio that could be utilized as the control signal in myoelectric prostheses or electrical stimulators. The advantage of using a programmable circuit was the capability to change circuit properties (cut-off frequencies of filter, gain and threshold voltage, etc.) by software during circuit performance. However, the system's high-power consumption restricted the use of batteries required for galvanic isolation.

In 2008, Cipriani *et al.* [33] designed three different hierarchical control methods using myoelectric commands, position and force sensors, and a vibrotactile feedback system. The prosthetic hand control strategy included two main parts: a low-level control (LLC) and a high-level control (HLC). The LLC loop was dependent on the sensory system and was used for grasp stability, whereas the HLC was adopted to provide appropriate signals for afferent stimulation and identify the subject’s purposes for executing desired action patterns. A finite-state machine was implemented to control prosthesis operation. The transitions between various states were detected as events by the LLC (automatic control) or by the HLC (interactive control based on user intentions). The differences between the three control strategies (M1, M2, and M3) were managing the grasping parameters by LLC or HLC. In M1, all of the grasping parameters were managed by the hand LLC. In the second control strategy (M2), the LLC managed Stability and pre-shaping, but the HLC managed the closure force. In M3, the LLC managed the stability, whereas the HLC managed the forced closure and pre-shaping. These control strategies were evaluated by experimenting on 14 able-bodied subjects. According to the results, the highest grasp success percentage (grasps were considered successful when the objects were correctly lifted off without slippage) for M2 and M3 was 90%. However, most users prefer M2 and M1 since M3 required a significantly higher level of attention.

In 2009, Zhang *et al.* [26] designed a control procedure based on the finite-state machine to control a meal assistance robot utilizing restricted muscle activities. They used EMG obtained from lower-limbs to control a robot, which contained an operating interface, a 5-DOF manipulator arm, and a 1-DOF end-effector. This system included four parts, an EMG amplifier, a detection part, a control circuit, and a meal assistance robot. The detection part employed an algorithm with an adaptive EMG power threshold to distinguish the start of muscle contractions in EMG signals. Three steps were required for the robot to do eating function using EMG signal: initial testing, adjusting the position of the end-effector, and selection of tray. In this study, states indicate predefined motion command for the robot, and FSM was applied to accomplish the transition of nine motion states, according to predefined muscle contraction (these states were achieved according to single click, double click, or no contraction of each calf). The authors compared the performance of the EMG system with the usual joystick system. The study results on four subjects showed that the average accuracy of EMG and joystick control systems was almost equal (92.1±5.6% vs. 93.6±6.6%, respectively). However, according to a two-way ANOVA test, the EMG control method had significantly better performance in fatigue, humanity, and flexibility.

In 2010, Došen *et al.* [34] developed an autonomic controller that included a vision system called the cognitive vision system (CVS) and a myoelectric control. The control algorithm contained four functions: collecting the inputs (image and distance from the CVS, EMG signals from the amplifiers), processing the data, creating the hand control commands, transferring commands to the hand. This control system employed a finite state machine to accomplish the shifts between the primary states (hand open and close) determined by the user's EMG. The FSM procedure is shown in Figure 5. The finger extensors (Ext EMG) and flexors (Flex EMG) muscles provided a two-bit binary code to drive the system through the states. The first bit of the code indicated the flexor's movement, and the second showed the extensors' function, while X meant "don't care." The user pointed the hand toward a purpose object and performed the hand opening. The CVS calculated the grasp type and size. The user touched the object, sent instructions to the hand to close, handled the object, and ultimately instructed the hand to open and free the object. The control system experimented on 13 able-bodied subjects. The results showed that the success rate in determining grasp type and size was 84%. Also, it was demonstrated that reducing the number of feasible commands improved classification accuracy. For example, accuracy was 93% for predicting the grasp alone.

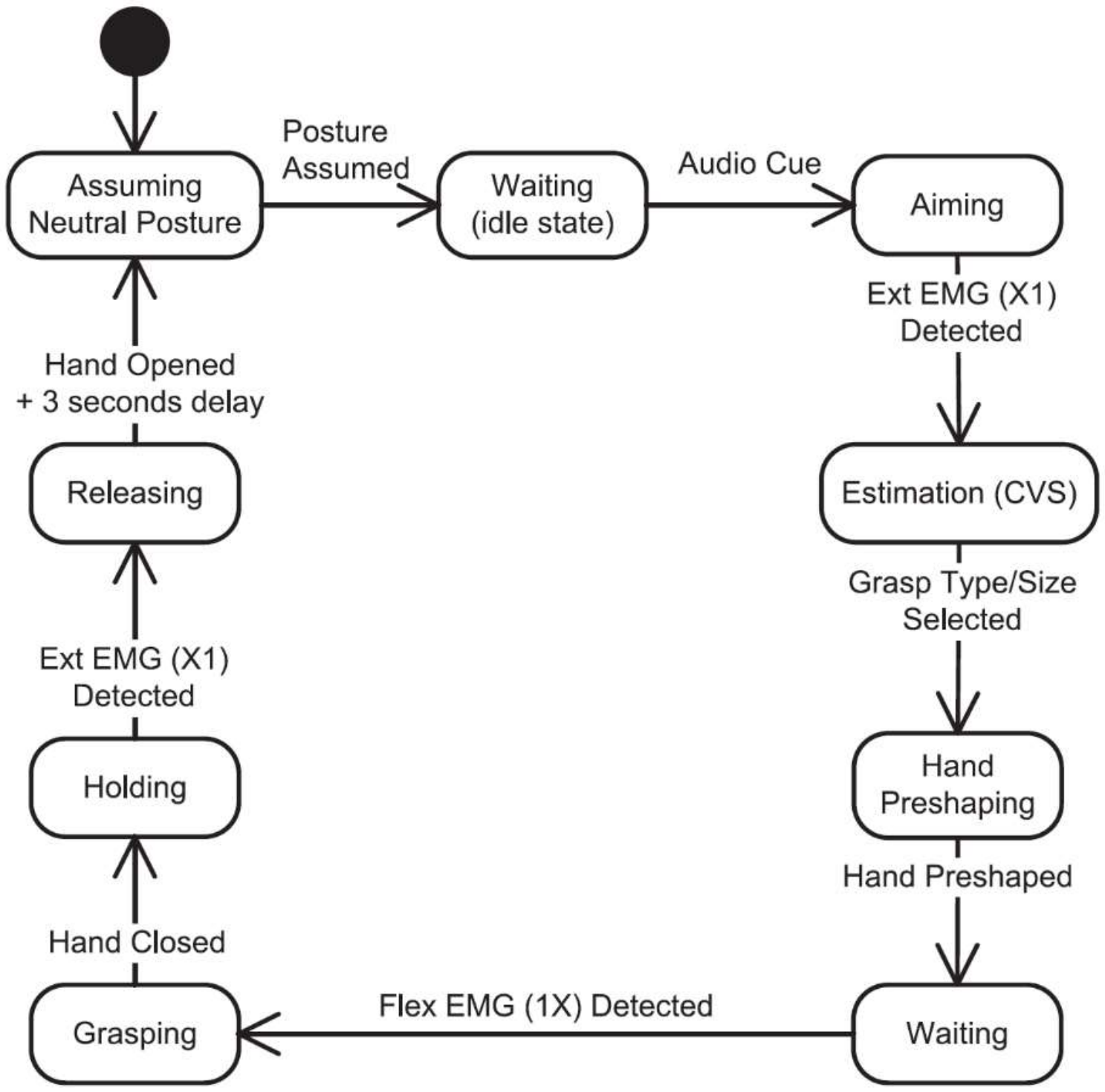

**Figure 5.** Finite state machine for autonomic controller developed by Došen et al. [34]. (Reproduced with permission from [34]).

In 2017, Krasoulis *et al.* [35] used a multi-modal approach combining sEMG with inertial measurements and a training data acquisition model to improve hand movements' classification performance. They applied a finite-state machine to accomplish real-time control of the prosthesis. A predicted movement by the classifier only was executed if the recently performed movement had finished accomplishment. For defining the accomplishment of movement, the motor's current data of the prosthesis were regularly observed and compared to a specified threshold. A control command was triggered barely when the last probability of the identical class surpassed the threshold. They examined the real-time control with 11 able-bodied and one amputee subject. The results indicated that this control method improved the classification accuracy with an 83% completion rate and 40 seconds of completion time.

In 2017, Controzzi *et al.* [36] designed the SSSA-MyHand prosthesis using an embedded controller combined with position and force sensors, which performed automatic grasps with only three electric motors. The controller had a master-slave structure with a pair of microcontrollers. The EMG signals were applied to the master microcontroller, which communicated with the external devices via the serial port and the bus of digital inputs/outputs, executed the finite-state machine that managed the hand operation, and controlled the actuators using torque, speed, and position PID algorithms. The slave microcontroller was responsible for computing the motors' actual position and speed, then passed this data to the master controller. The primary states were the pre-shape postures that were related to five possible functions by the hand. These states were sequentially selected using EMG co-contractions. Once in a pre-shape state, the hand could enter a grasping state and be managed by a two-state amplitude-modulated myoelectric control method with noise thresholds. In a pre-shape state, the prosthesis could return in a rest position if the EMG activity was less than the threshold for a specified time.

In 2018, Bahrami Moqadam *et al.* [21] used the combination of on-off myoelectric control with a fuzzy/PD method to provide grip position and grip force controlling of the prosthesis. In the first step of the external control loop, a two-channel EMG signal was recorded. Then the next block improved this signal and reduced its noise range. In the next phase, the EMG signal was rectified and smoothed using a low-pass filter before passing through the analog to digital converter. In the logic circuits, two thresholds were set to determine hand movement's direction (clockwise or counterclockwise). This loop's output was similar to a stop signal, which was the inner control loop's input. In other words, the voltage domain of the EMG signal affected the output of the fuzzy logic section, and the value of the time difference produced the input of the PD controller (Figure 6).

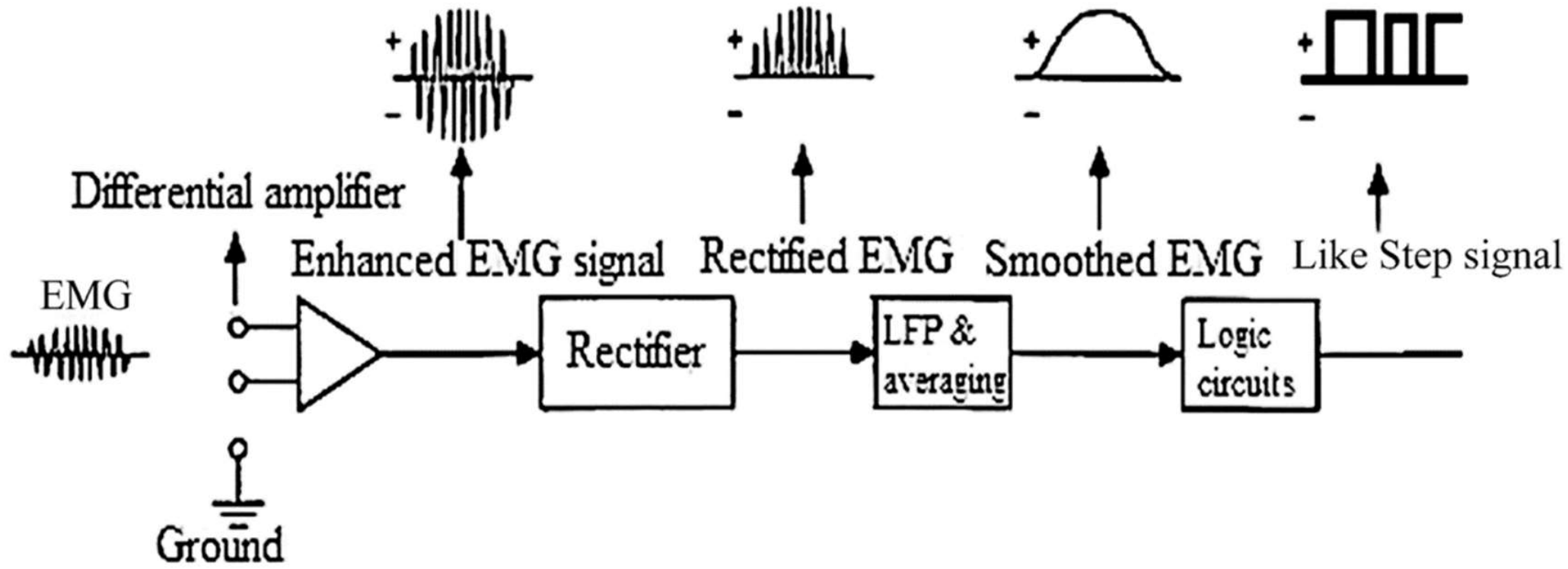

**Figure 6.** Block diagram of the myoelectric control part (external loop) developed by Bahrami Moqadam *et al.* [21]. (Reproduced with permission from [21])

In 2019, Leone *et al.* [37] developed a hierarchical classification procedure to determine both intended hand/wrist postures and desired force levels during grasping functions. They utilized a finite state machine to control and organize three classifiers based on NLR (Non-Linear Logistic Regression algorithm). The FSM strategy could only be in a limited collection of states and one state at a time and receive a series of inputs. The highest NLR classifier, the "hand/wrist gestures classifier," was applied to distinguish the intended gesture class between seven different gestures. The next classifier in the hierarchy was defined according to the output of the first classifier. If the output of the highest classifier was "Spherical" motion class, the "Spherical force classifier," and if it was "Tip" motion class, the "Tip force classifier" was used to specify the intended force level among three different force levels (Low, Medium, and High). Otherwise, if the output of the highest classifier were something other than "Spherical" or "Tip" among hand/wrist gestures, the force classifiers would not be activated (Figure 7). The application of FSM provided the feasibility of simultaneous operation of two classifiers of different levels of the hierarchy (Figure 8). This procedure was evaluated on 31 able-bodied subjects who were asked to perform seven hand/wrist movements and apply three different force levels during the "Spherical" and "Tip" grasps. The results showed that the NLR classifiers presented great accuracy for gestures (98.78%) and forces (Spherical 98.80% and Tip 96.09%). These results were compared to benchmark LDA classifiers results (gestures 95.41%, Spherical 98.74%, and Tip 97.60%). According to a Wilcoxon Signed-Rank test, no statistically significant differences in F1Score performance were observed between NLR and LDA.

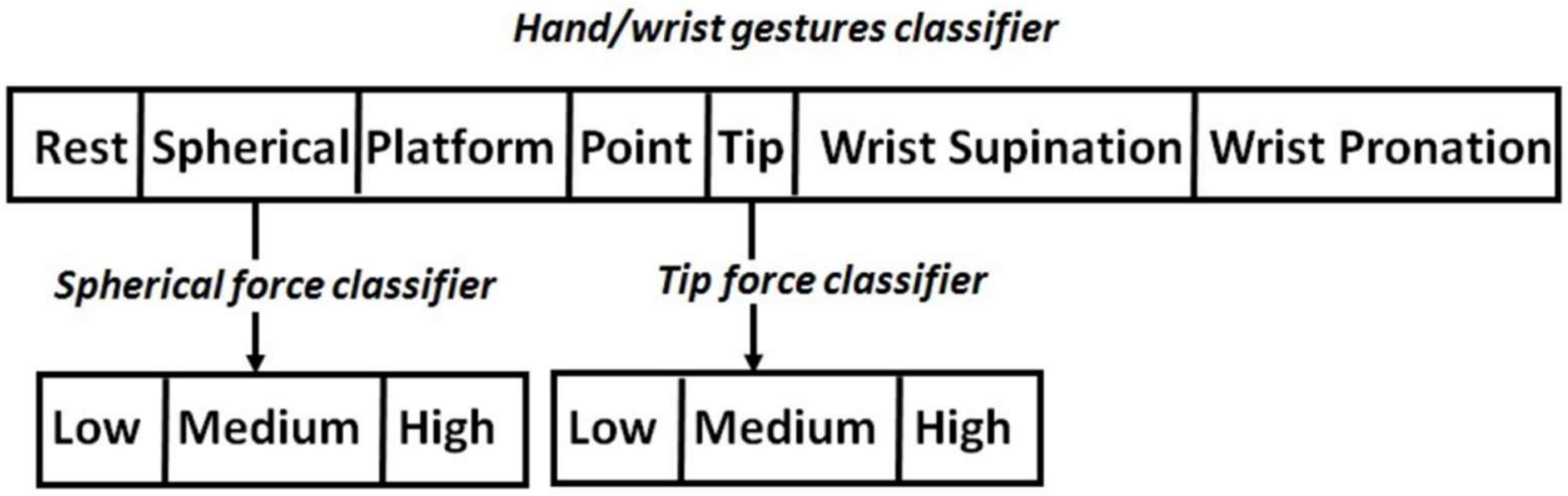

**Figure 7.** Hierarchical classification of hand/wrist gestures and force level. (Reproduced with permission from [37])

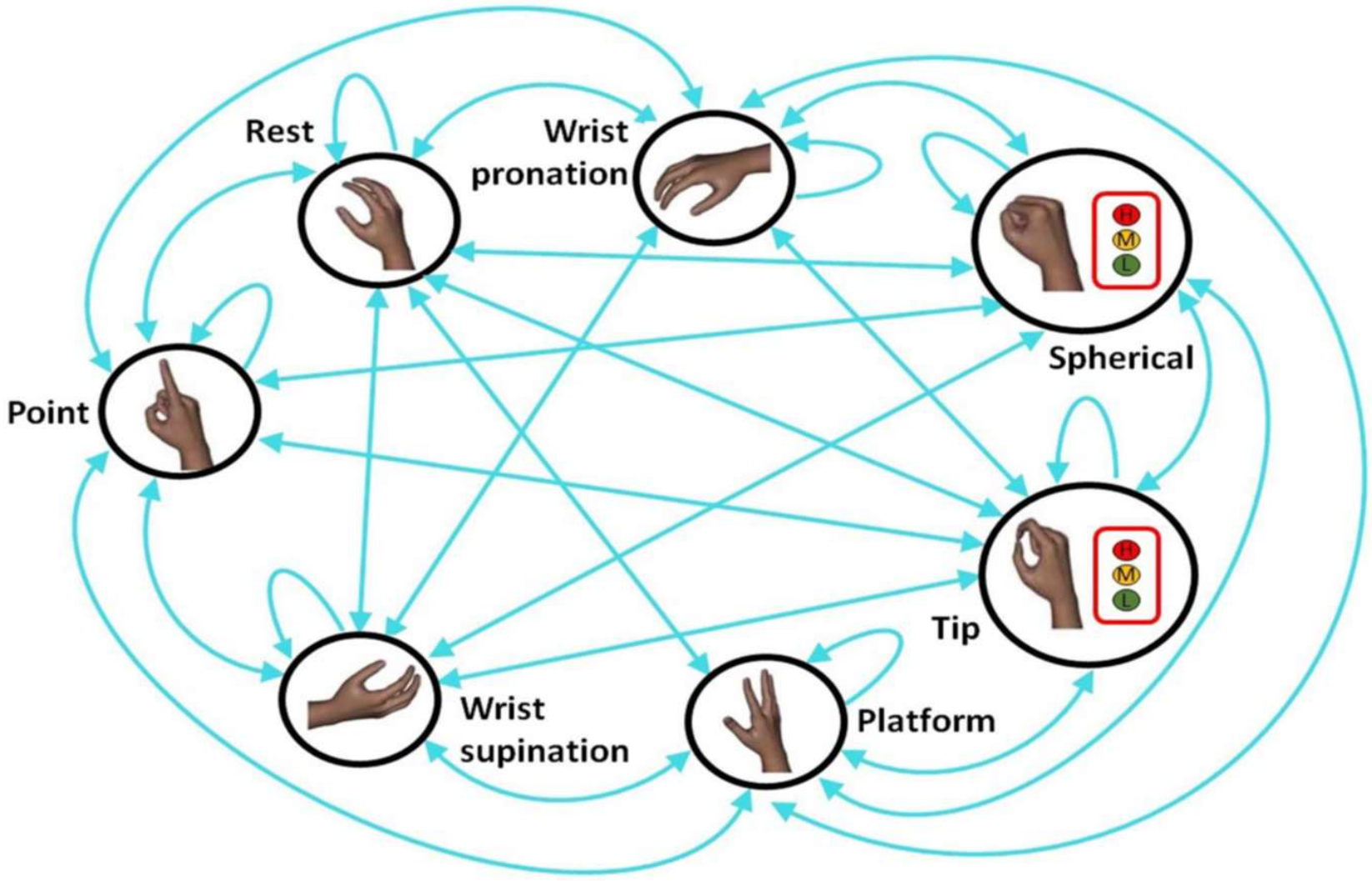

**Figure 8.** Finite State Machine approach for the classification of seven different hand/wrist gestures and three force levels. (Reproduced with permission from [37])

### *3.2. Proportional, direct, and myoelectric posture control*

In proportional myoelectric control, the motor voltage varies in direct proportion to the EMG signal. It thus gives the amputee control over speed and force of grip [38]. In 1991, Sears and Shaperman compared on-off control with a simple proportional myoelectric control. The experiment was conducted on 33 patients dividing into different groups based on their experience with a prosthesis. Former on-off controlled prosthesis users rated the proportionally controlled hand higher in aspects like quickness, speed and force control, and required effort for opening and closing hand [38].

Feature extraction or parameter estimation has a significant role in proportional control. The first Proportional myoelectric controlled prosthesis was introduced by Bettye *et al.* in 1955 [27]. The constructed apparatus was sensitive to a light grasp, and the user could close the grip by contracting the muscles and maintain it with less effort. The proposed system was not practical for daily life, and reliability was the main focus [27]. In 1963, Bottomley *et al.* proposed a system for proportional control of force and speed from surface EMG (sEMG) channels, using mean absolute value (MAV). The output voltage was proportional to the differential signal from biceps and triceps muscles. The remaining fluctuations after rectifying and smoothing decreased by using the backlash circuit [27,38–40]. Myo-pulse (MYOP) is another feature that was used by Philipson in 1985 [41]. In this method, similar to pulse width modulation (PWM) control, the value of the output voltage is "1" when the EMG signal is greater than the threshold; otherwise, the output is "0".

Moreover, the duty cycle of the output voltage is proportional to EMG activity (Figure 9). This feature can be obtained by using only a dual comparator after amplification [41]. In 1977, Parker *et al.* proposed a formulation based on the detection problem, which resulted in optimal parameter values and signal processing scheme [42]. Hogan *et al.* 1980 suggested a similar mathematical statement for an optimal myoelectric signal processor. This feature was also known as root mean square (RMS) [43]. In 1984 Evans *et al.* suggested applying logarithmic nonlinearity and a linear minimum mean-square error estimator instead of a full-wave rectifier and a low-pass filter. The Kalman filter was then used to estimate the control signal [44].

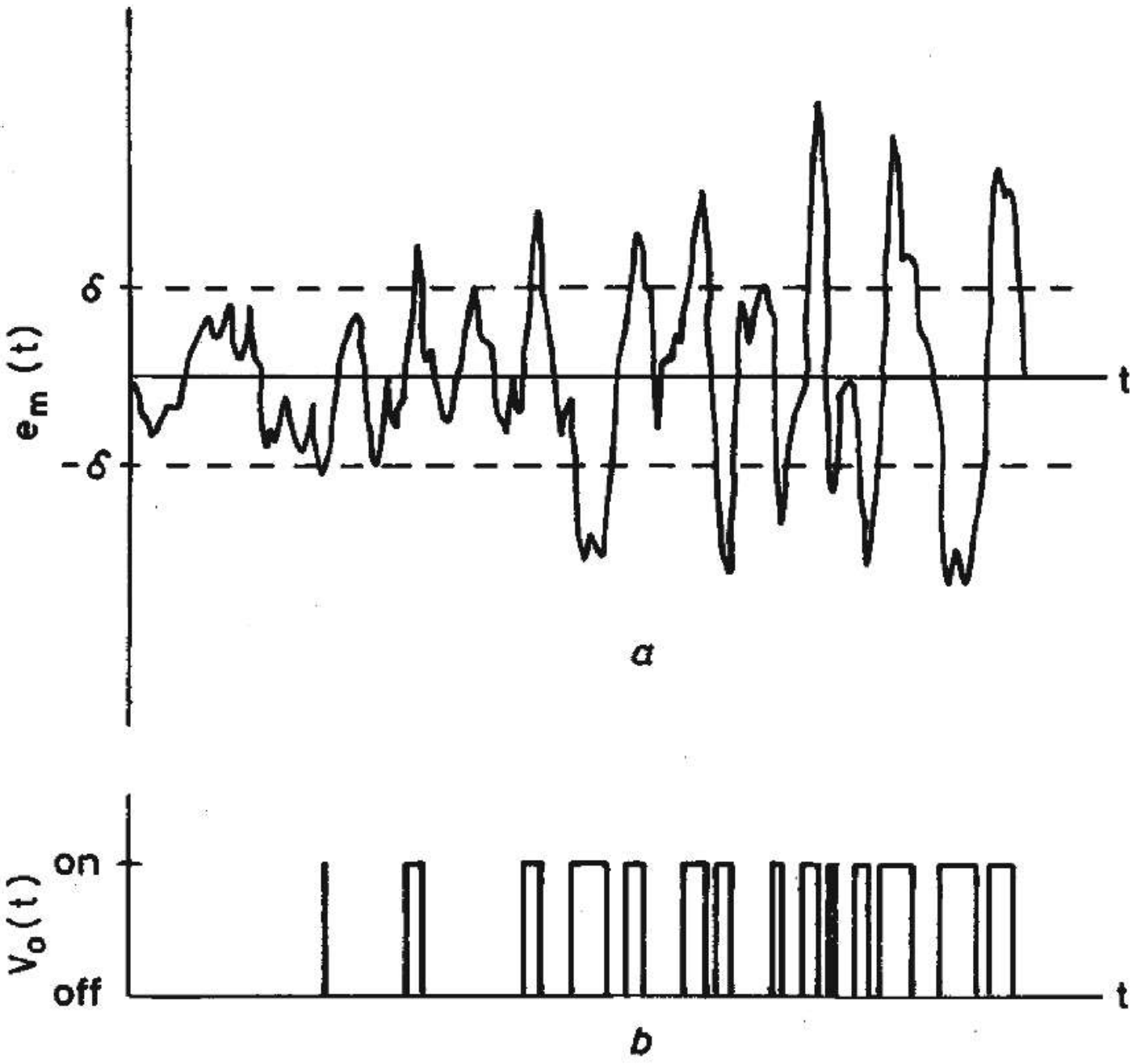

**Figure 9.** Illustration of the myopulse (MYOP) processing technique. A typical bandpass-filtered sEMG signal is shown in the upper trace (a). The lower trace (b) illustrates the output from the myopulse processor. The output is turned on when the absolute amplitude of the sEMG signal is higher than a threshold $\delta$. Otherwise, the output remains low. (Reproduced with permission from [41]).

In 1993, Hudgins *et al.* [45] developed a popular set of time-domain (TD) features (slope sign changes (SSC), waveform length (WL), mean absolute value slope, MAV, and ZC) for controlling a multifunction prosthesis by a single EMG channel, without requiring extra effort from the user. The artificial neural network (ANN) was used for classification. Results demonstrate that the overall accuracy for nine healthy subjects was 91.2% (SD 5.6%) and for six amputees was 85.5% (SD 9.8%). Englehart *et al.* [46] demonstrated that short-time Fourier transform (STFT), the wavelet transforms (WT), and the wavelet packet transform (WPT) outperformed TD features and provided better performance for classification. The average classification error for the suggested features was 6.25% (or 93.75% accuracy), and for Hudgins' TD features average error was 9.25% in 16 healthy subjects. In 2009, Jiang *et al.* [47] suggested multiple degrees of freedom (DOF) activation using nonnegative matrix factorization (NMF) algorithm and the mean square value (MSV) of the multichannel sEMG. Twelve intact subjects participated in the experiment. The results demonstrated that force estimation was successful using MSV for simultaneous and proportional control.

Many of the recent studies have also focused on proportional myoelectric control. For instance, in 2012, Muceli *et al.* [48] suggested a method for simultaneous and proportional control at four DOF. Six healthy subjects performed flexion/extension, radial/ulnar deviation, forearm pronation/supination, and hand closing. The accuracy of kinematic estimation from sEMG was in the range of 79%-88%. In 2014, Fougner *et al*. [49] proposed simultaneous and proportional control with system training based on prosthesis-guided training (Figure 10). The study was conducted on a prosthesis with two motor functions. The EMG features were average amplitude value (AAV), ZC, WL, and MYOP. Two normally-limbed subjects participated in the study. Compared to the other three control schemes performed in this study (mutex proportional control, mutex on-off control, and sequential proportional control), simultaneous and proportional control was similar to two mutex-based controls outperformed sequential proportional control.

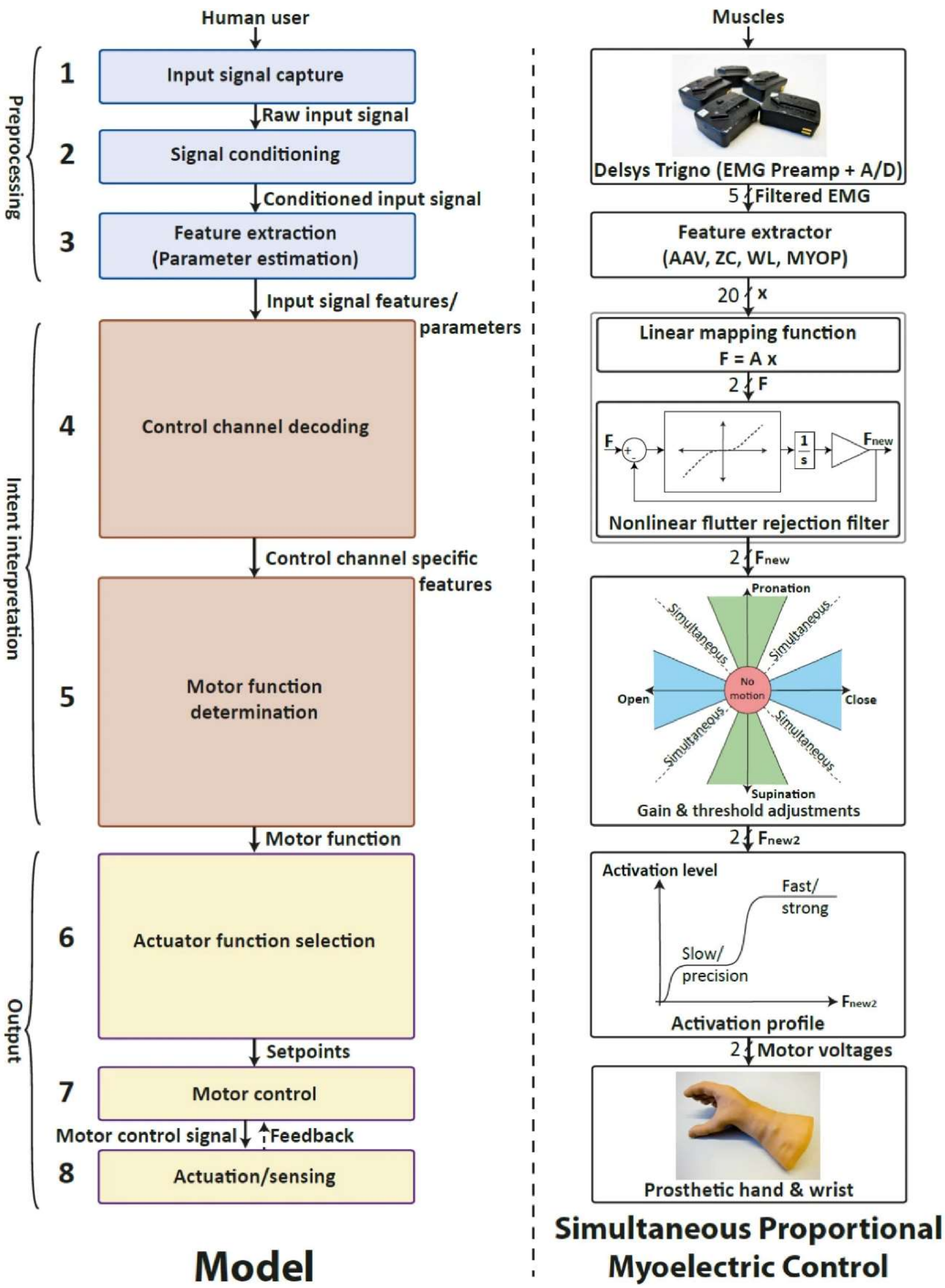

**Figure 10.** An overview of the control structure used for simultaneous proportional control. (Reproduced with permission from [49]).

In 2014, Tang *et al*. [50] developed a proportional myoelectric control for an upper-limb power-assist exoskeleton in real-time. Six able-bodied subjects performed elbow movement (flexion/extension) during four experiments with different conditions. sEMG data were recorded using four pairs of electrodes, and RMS was used as the feature for the back-propagation neural network (BPN) classifier (Figure 11). The RMSE and R-square showed the error (fitness) between the actual and the predicted angle. The four seconds results had a lower RMSE (9.67) and greater R2 (0.87) values.

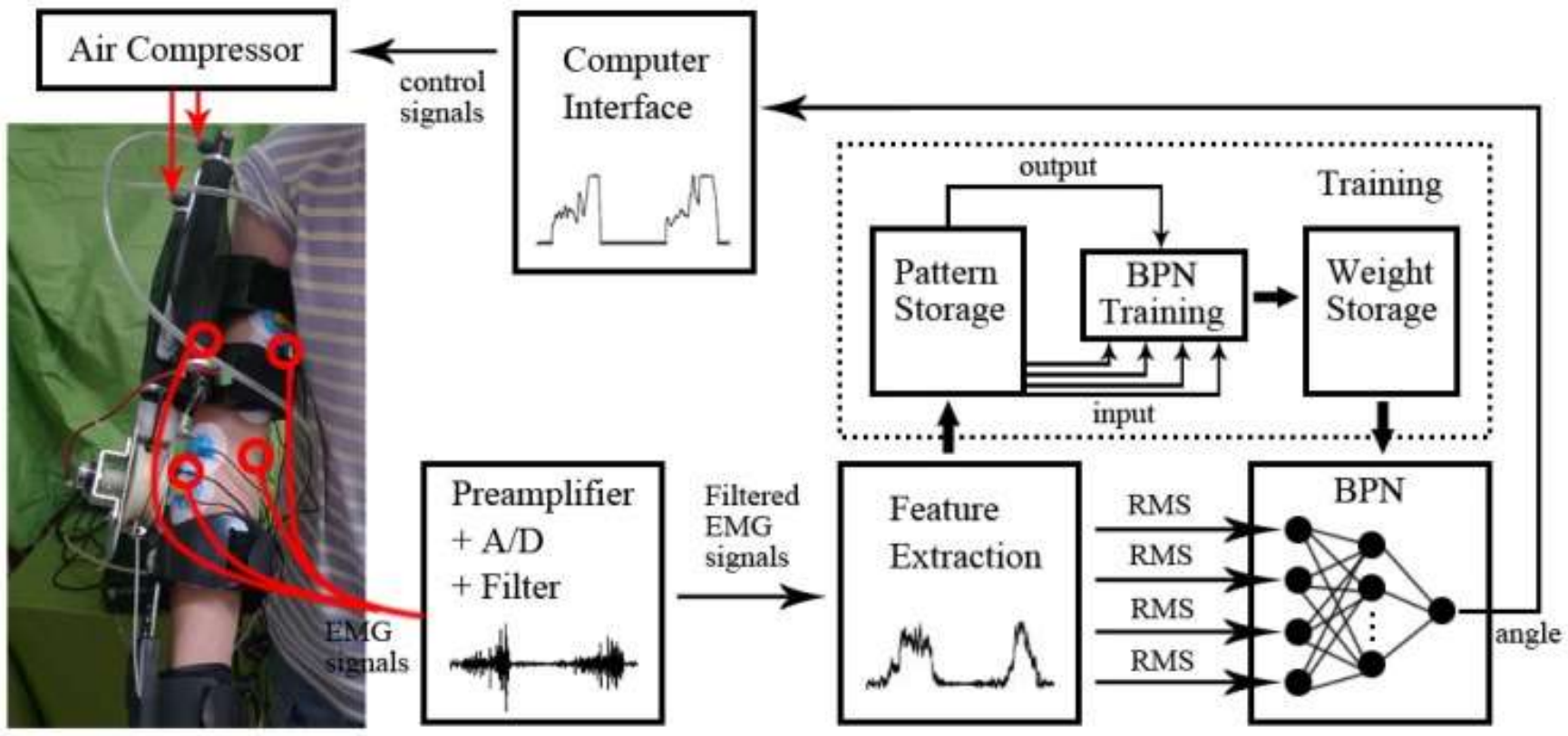

**Figure 11**. The proposed proportional myoelectric control scheme (BPN: back-propagation neural network, EMG: electromyogram, RMS: root mean square) (Reproduced with permission from [50]).

In 2014, Jiang *et al.* [51] demonstrated an online simultaneous and proportional control based on NMF. Seven subjects with unilateral limb deficiencies and seven intact subjects participated in the experiment. They were able to perform two wrist DOF (flexion/extension and pronation/supination). The results showed that both intact and amputee subjects' performance was similar (the average completion rate was 95%). They also presented a state-based proportional myoelectric control, which provided natural and intuitive switching between functions [52]. Nine healthy subjects and one subject with limb deficiency participated. The average completion rates for 1-DOF and 2-DOF tasks and grasping force were, respectively, 96.2%, 91.5%, and >97%. In 2015, Amsuess *et al.* [53] proposed a proportional speed control based on the common spatial patterns (CSP). It outperformed a pattern recognition method, linear discriminant analysis (LDA), and academic state-of-art methods. Tests were performed on four amputees and ten able-bodied subjects.

Simultaneous, independent, and proportional control of a hand prosthesis with multiple DOF and fine movements is possible by mapping each EMG channel directly to a specific function. This method is known as direct control (DC). The number of muscle sites for electrodes is limited; therefore, this control strategy requires mode switching for more than one DOF performance [54,55].

Many studies have compared DC with other myoelectric control strategies. For instance, in 2014, Wurth *et al.* [55] compared three control strategies: DC, sequential pattern recognition (PR), and simultaneous PR in real-time. Nine intact subjects and two amputees participated in this study. Tests were based on the principles of Fitts' law. Simultaneous PR's performance in throughputs and path efficiencies was significantly better than the other two strategies (p<0.01). In 2016, Kuiken *et al.* [56] compared DC with PR. In DC, the user must switch modes for using wrist motor or selecting different hand grasps, which creates slow and demanding control, while a little cognitive effort is required. Three amputees who were familiar with the DC method participated in the home trial using PR. Classification error rates for two of the three subjects were obtained (Error rates before home trial were 9.3% and 16.4%, and after the home trial, error rates were 12.9% and 10.2%). Results showed that subjects could perform wrist rotation, wrist flexion, and two hand grasps using PR. In 2017, Hargrove *et al.* [57] similarly compared PR and DC. Eight amputees who have had targeted muscle reinnervation (TMR) participated in a home trial using DC and PR. PR outperformed DC in the Southampton hand assessment procedure (p=0.041) and the Clothespin relocation task (p=0.024). These experiments include movements with three DOF. In another study in 2018, Ransik *et al*. [58] compared self-reports and PR and DC performance for a two DOF prosthesis controlled by a transradial amputee. Outcomes from both control strategies were similar in 75% of the metrics.

One of the challenges in using direct control is crosstalk [59] in sEMG signals, decreasing sensitivity and efficiency; therefore, intramuscular EMG (iEMG) recording has been suggested. Using iEMG, noise from neighboring muscles can be reduced, and obtaining signals from deep muscles is possible.

In 2009, Kamavuako *et al.* [60] quantified the correlation between grasping force and the recoding methods (sEMG and iEMG) from ten healthy subjects. Results showed that the correlation for iEMG and sEMG were similar (~0.9), and iEMG could be used for proportional control. In 2014, Smith *et al.* [61] proposed a simultaneous and proportional myoelectric control of three DOF (wrist rotation, wrist flexion/extension, and hand open/close) in real-time on five intact subjects using iEMG. The results showed that the parallel dual-site differential control, which provides simultaneous prosthesis control, outperforms pattern recognition. Similarly, in 2018, Frost *et al.* [62] acquired iEMG signals from a neural interfacing method called regenerative peripheral nerve interfaces (RPNI) in a rat model for real-time and proportional control of prosthesis hand (Figure 12). This method was not affected by motion artifact or crosstalk from neighboring muscles.

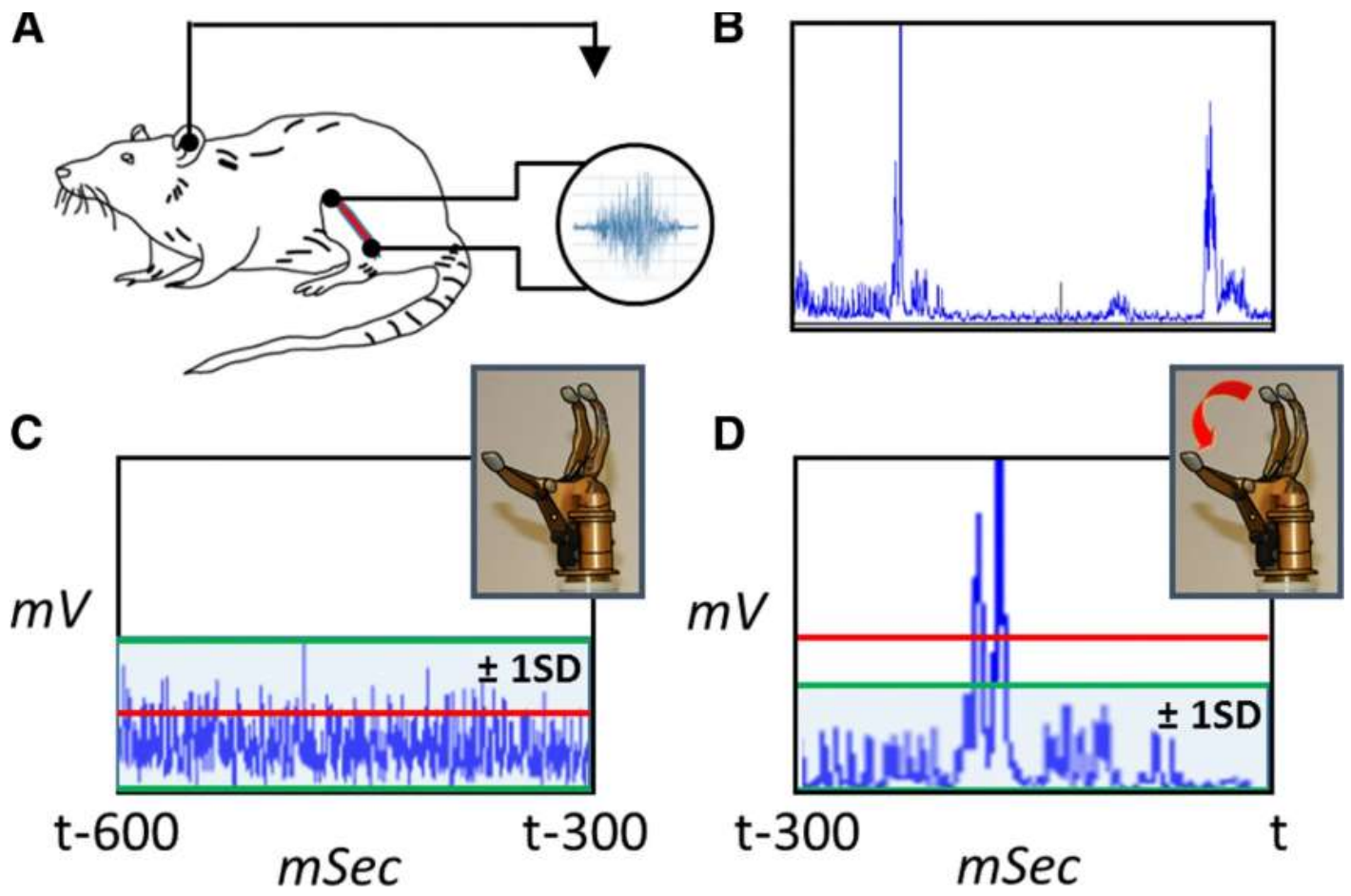

**Figure 12.** The activation process of prosthesis hand, proportional to iEMG signal from an RPNI rat. A. Acquisition of raw iEMG signal. B. Filtering and rectifying the recorded signal. 300 msec interval containing C. no leg movement and D. leg movement, which shows hand activation when the real-time iEMG is greater than the threshold (EMG signal; Red lines: iEMG value; Green lines: Activation threshold) (Reproduced with permission from. [62]).

Another strategy for simultaneous myoelectric control of prosthesis is posture control. Due to the lack of EMG channels, the ability to continuously transform between postures is decreased (multiple-input multiple-output (MIMO) problem) [63]. In 1998, Santello *et al.* [64] investigated the grasp postures of 57 objects. The study showed that the first two principal components accounted for more than 80% of the variance, implying a substantial reduction from the inputs. In 2010, Mtrone *et al.* [65] developed a control algorithm using principal component analysis (PCA) to control a 16 DOF prosthesis (CyberHand). Later, in 2012, a myoelectric controller based on a PCA algorithm was designed [66]. Two differential sEMG channels were collected as input data. Twelve healthy subjects performed several tasks (grasping, transporting, and releasing), and simultaneous control with two-DOF wrist movements was utilized (Figure 13). The CyberHand postures distribution over the two-channel input signals plane was shown in Figure 14.

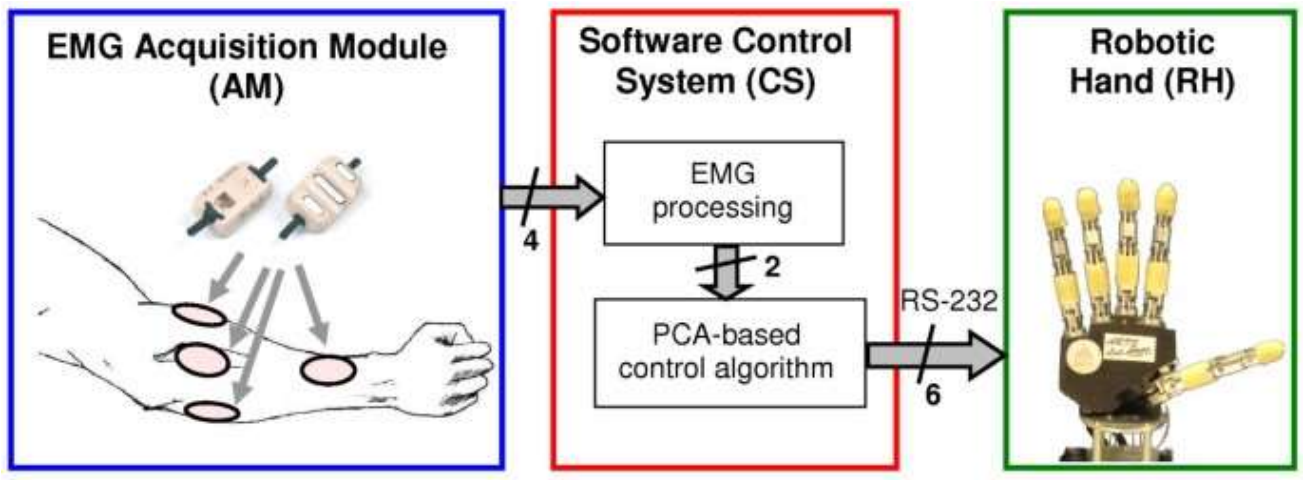

**Figure 13.** Overview of the proposed system for myoelectric control of a hand prosthesis using PCA (principal component analysis). (Reproduced with permission from [66]).

Wrist flexion/extension (Ch1) controlled fingers flexion/extension and wrist adduction/abduction (Ch2) controlled thumb rotation (Figure 17 and 18) [66].

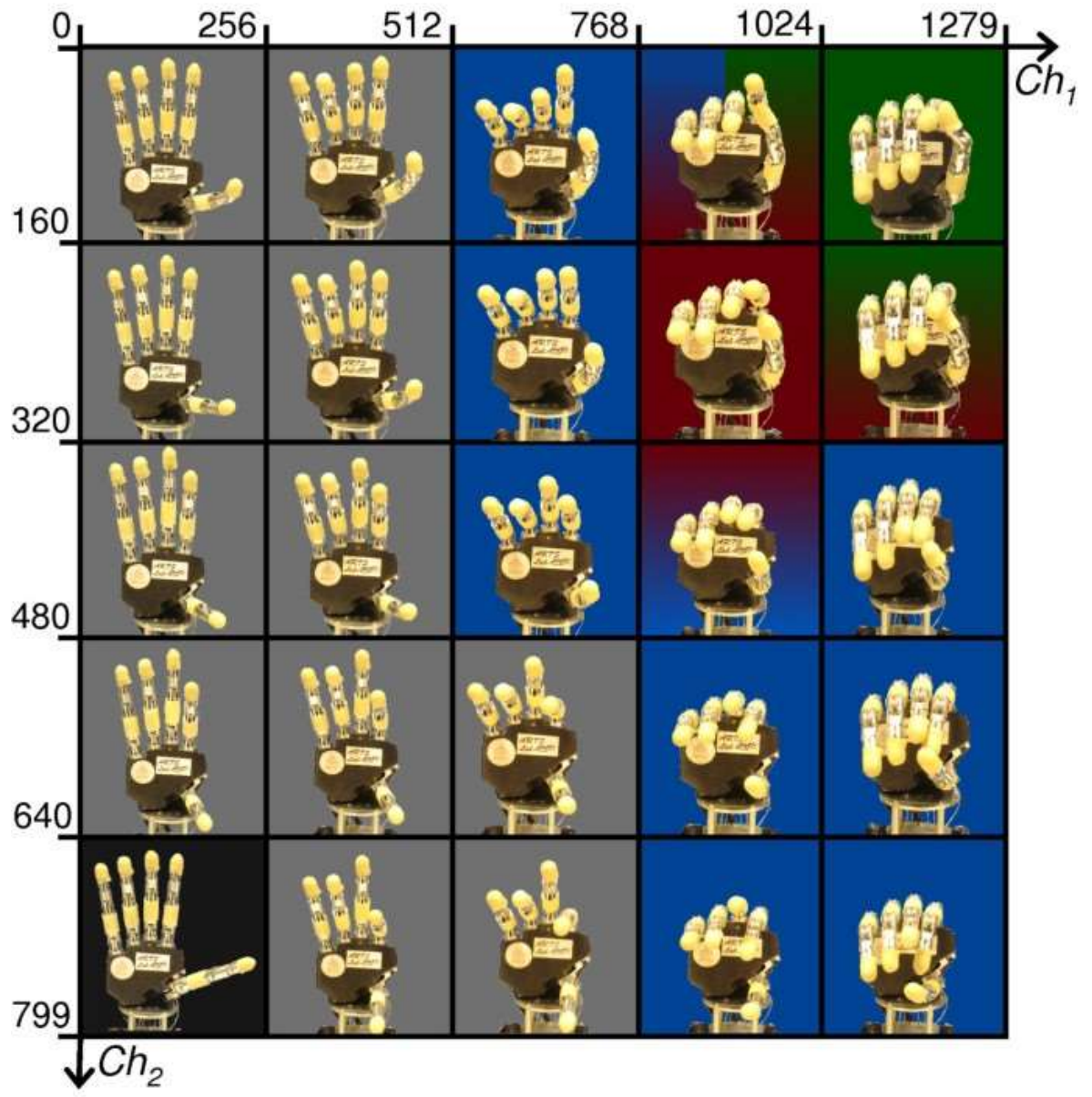

**Figure 14**. Postural behavior grid. Blue, red, or dark green colors indicate respectively power, precision, or lateral grasp. Faded backgrounds indicate areas with the possibility of multiple grasp types. The black color indicates the open-hand neutral posture. (Reproduced with permission from [66]).

In 2014, Segil *et al.* [67] also used postural myoelectric control using PCA. Ten amputees participated in this study. Four EMG signals were converted into a principal component domain using four maps. These maps controlled 11,13,10 and 11 out of 15 joint angles accurately. Xu *et al.* [68] similarly used PCA to obtain the first two postural synergies and control 12 output tendons of an 18-DOF prosthetic hand. Two-dimensional control signals were obtained from six sEMG channels. Four intact subjects and one amputee participated in this study. Results indicated that all subjects were able to control the prosthetic hand using two postural synergies. In 2015, Segil *et al*. [69] used a postural control system tested by eleven subjects. The controller mapped EMG signals into a joint angle array. Vector summation algorithm (VSA) was implemented on the sEMG RMS signal. The completion rate (CR) of velocity cursor-control (84±3%) was greater than that of the position cursor-control (45±3%). The number of electrodes (i.e., 3,4, or 12) did not affect the results.

In 2017, Segil *et al.* [70] examined the postural control system using a Bebionic hand. Four subjects with transradial limb loss and four healthy subjects participated in this experiment. Subjects performed the Southampton hand assessment procedure (SHAP), which included picking up coins and rotating the key and door handle. All subjects could perform the tasks successfully. Results showed that subjects with limb loss restored 55% of normal hand function.

### *3.3. Simultaneous myoelectric control*

Despite the advances of recent decades in myoelectric control, there is a huge gap between industrial and academic prostheses. In general, these prostheses have a long way to go until they reach the natural hand, so most users do not accept these prostheses and prefer to cope with their disabilities. One of the most important reasons for users' dissatisfaction is the lack of simultaneous control. It is necessary to activate several degrees of freedom (DOFs) simultaneously in most daily life movements. However, these DOFs are activated sequentially in industrial prostheses, so these prostheses' movements are not smooth and do not act like dexterous hands. For this reason, more attention has been paid to simultaneous control in recent years. Many methods have been proposed in the literature in this area. These methods include pattern recognition, deep learning, regression, and non-negative matrix factorization (NMF) [23,71,72]. In the following, some of these studies are discussed.

In 1973, Herberts et al. proposed the idea of simultaneous control for the first time [73]. They tried to classify seven hand movements (3 DOFs) using an artificial neural network (ANN). No quantitative results were reported for this study, but they reported that it was possible to control several DOFs simultaneously.

So far, different strategies have been used for simultaneous control based on pattern recognition. The most straightforward strategy is adding combination movements as separate classes [74–76]. For example, in [77], LDA was used in this approach. The results of this study showed that this method could control several DOFs simultaneously. Another strategy is using multiple binary classifiers, a.k.a. the one-vs-one method. In this approach, several classifiers are trained to discriminate between two classes of motion, and majority voting is used to make the final decision [74,75]. Scheme *et al*. [78] considered the LDA classifier for applying this method. The results showed less classification error than previous approaches and other classifications such as KNN. Another strategy uses multiple classifiers to distinguish between one class and combining all other classes, a.k.a. one-vs-all. This method was used in [79] with LDA and in [80] with the SVM classifier. A more complex strategy applied in some studies is the hierarchical method [74]. Young *et al.* [81] introduced a hierarchy of LDA classifiers. The results of this study showed a classification error of less than 15%.

Pattern recognition-based models detect discrete movements. Thus, in some studies, pattern recognition-based regression methods have been used to detect movements continuously. These methods allow simultaneous and proportional control [72]. For instance, the bootstrap aggregating (bagged) regression trees method was used for the first time in the field of simultaneous control by Ameri *et al.* [82]. In this study, three DOFs were considered. This study showed that compared to ANN, the proposed method is better in the abduction-adduction DOF, and there is no significant difference between the two methods in other DOFs. Two methods were compared in terms of processing time, which showed that the proposed method's processing time was less than ANN. Due to its short processing time and almost the same function as ANN, this method could be used as simultaneous control.

Following the previous work, Ameri *et al.* [83] evaluated the support vector regression (SVR) performance in the field of simultaneous control. Data were recorded from 10 healthy subjects and two amputees. The results of this method were compared with those of ANN. Evaluations showed that the SVR method is better for healthy subjects, but there is no significant difference between the two approaches for amputees. Since the number of DOFs was higher in healthy subjects (3 DOFs in healthy subjects and 2 DOFs in amputations), it can be concluded that SVR outperformed in higher DOFs. Ameri *et al.* [84] used CNN for simultaneous control in 2018. They compared the results of this study with SVM. The results of this study showed that there is no significant difference between CNN and SVM.

In 2020, Chenfei *et al.* [85] proposed a short connected autoencoder long short-term memory (LSTM) to estimate continuous arm movements. Two current models (CNN and MLP) were used to compare the results. This experiment showed that the performance of the proposed method was significantly higher than CNN and MLP models. The proposed method's overall accuracy, CNN, and MLP were 95.7%, 86.8%, and 83.4%, respectively. Bao *et al*. [72] combined two deep learning methods of CNN and LSTM. In this experiment, CNN was first used to extract the feature from sEMG, and then the LSTM was used to estimate the wrist kinematics. The results showed that the combined method performed significantly better than CNN and LSTM individually, mainly in more complex movements.

.

In 2014, Hahne *et al*. [86] evaluated different types of linear and nonlinear regression methods in the field of simultaneous and proportional control. In this study, regression methods such as linear regression, Mixture of linear experts (ME), multi-layer perceptron (MLP), and Kernel ridge regression (KRR) were investigated. In this study, ten healthy subjects and one person with congenital impairment participated. The results of this study showed that KRR has better performance than other methods in most cases. On the other hand, it was shown that using transformation such as logarithmic transformation can make the relation between sEMG and wrist angle more linear, so linear regression methods that have less computational cost than KRR can be used. Due to the relatively good results of the ME method and its low computational cost, this method can be a suitable option for next-generation prostheses. One of the notable points of this study was that training data does not need to be collected for all modes. In other words, these methods can be generalized and can predict even states that do not exist in the training data if the number of training data is sufficient for each state.

In 2020, Piazza *et al.* [87] suggested the probability-weighted regression (PWR) method in the field of simultaneous and proportional control. This method was previously used for intramuscular EMG signals and showed a good performance. In this study, the performance of the PWR method was examined on the sEMG signals. Eight healthy subjects and four amputations participated in the study. The performance of this method was compared with that of the direct control. They found that for three DOFs in healthy subjects, the throughput was significantly better for the PWR method, but in the metrics of completion rate and path efficiency, the proposed method has lower performance than the direct control method. There was no significant difference between the two methods in amputation subjects for three DOFs, but in two DOFs, the proposed approach was better. In general, they concluded that using the PWR method, the performance was reduced for three DOFs, and for two DOFs, excellent results were obtained. One of the remarkable features of this research is investigating the proposed method's performance in daily tasks such as opening and closing the refrigerator door and drinking water from the bottle. All users mentioned that the PWR method had a closer functional to natural hands.

Another method used for simultaneous and proportional control is NMF [88]. The biologically generative model inspired this method that was first developed by Jiang et al. [47,89]. The proposed model diagram is presented in Figure 15, wherein F(t) is the force vector, S is the activation coefficients vector, X(t) is the activation function vector, M is the number of muscles involved in the movement, Y(t) is the sEMG produced before passing through the skin, G(t) is the tissue filter, and Z(t) is the recorded sEMG.

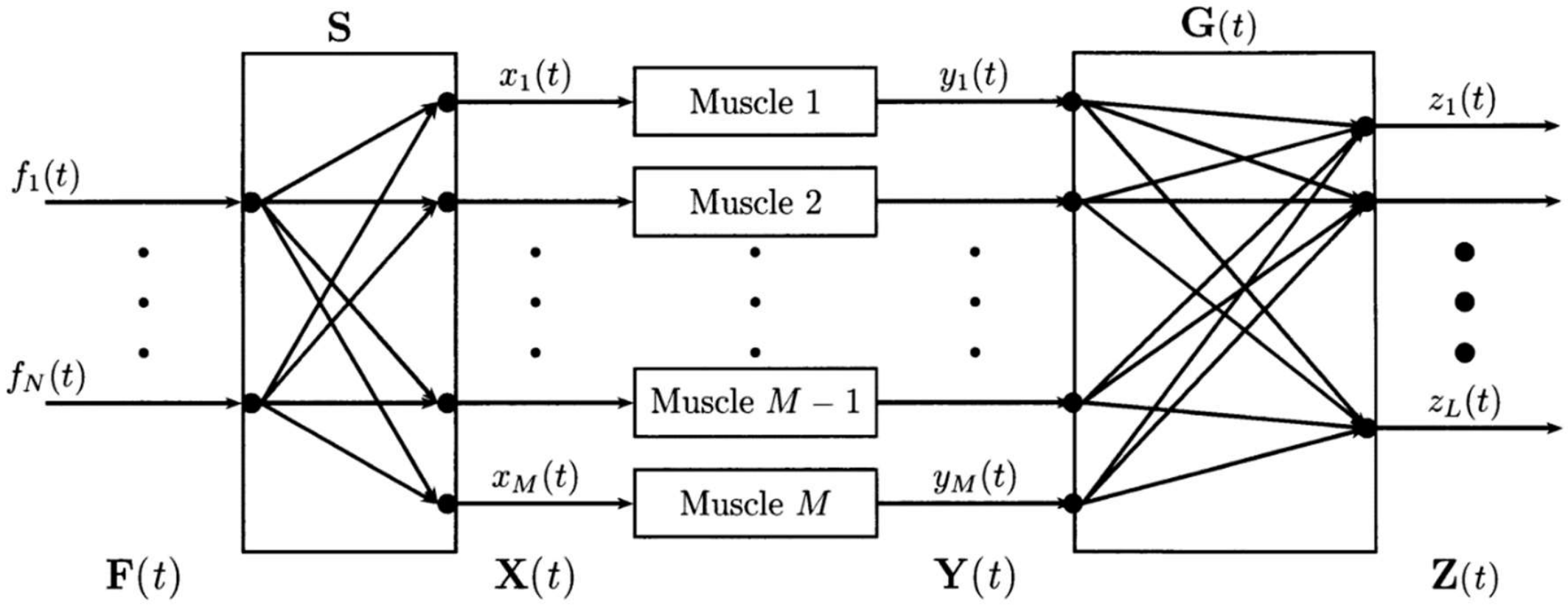

**Figure. 15.** Schematic of the proposed model for surface EMG. (Reproduced with permission from [89]).

They extracted mean square value (MSV) as a feature and assumed that the cross-talk between the channels was negligible. Therefore, the relation between MSV and the joint force is almost linear, and the relation can be expressed as:

$$Z(t) \approx W \cdot F(t) = \begin{bmatrix} W_1^+ & W_1^- & \dots & W_N^+ & W_N^- \end{bmatrix} \cdot \begin{bmatrix} F_1^+(t) \\ F_1^-(t) \\ \dots \\ F_N^+(t) \\ F_N^-(t) \end{bmatrix} \quad (1)$$

For each movement, $Z(t)_{M\times 1}$ is the feature vector (MSV) at time $t$, $W_{M\times 2N}$ is the muscle synergy matrix, and $F(t)_{2N\times 1}$ is the activation coefficient vector. Thus, the previous equation can be rewritten as:

$$Z(t)_{M\times L} = W_{M\times 2N} \,.\, F(t)_{2N\times L} \qquad (2)$$

where $M$ is the number of channels and $L$ is the signal length. In this scheme, simultaneous control of multiple DOF is obtained with the linear summation of single DOF activations. Therefore, during training, only single DOFs can be considered. The W vector for each motion is calculated using the NMF method to solve this equation in the calibration phase. The W matrix is created by putting the W vectors together (this matrix is considered constant). During the test, the pseudo inverse of this matrix is calculated, and by multiplying in the feature matrix, the activation coefficient matrix is obtained [90–92]. The $R^2$ index was used to evaluate the results, and for two DOF, it was 77.5+10.9%, but the results showed that the $R^2$ value decreased by adding the third DOF (pronation/supination). The critical point of this method is that it is a semi-unsupervised method [47]. Also, some studies have shown that this method is robust to the number of channels and the displacement of electrodes [93].

Nielsen *et al.* [94] used TD features instead of MSV, and the combination of DOFs is considered during training. They showed that by making these changes, the NMF method's performance could be increased by 38.9% and 14% for 3DOF and 2DOF, respectively. In another study, Ma *et al.* [92] investigated the effect of different NMF method performance features. They reported that the RMS, MSV, and WL performed well, while RMS and MSV perform better when the mean is zero. Concisely, it is better to remove the DC before extracting these features [90].

The most important limitation of the NMF method is its DOF-wise calibration phase, meaning that only single DOFs must be activated during training, while more than one DOF may be activated in the training phase. It leads to producing an inadequate synergy matrix and poor control [90,91]. To overcome this limitation, Lin *et al.* [24] proposed using the NMF method to include sparseness constraints. They selected the $L_1$-norm-based sparse non-negative matrix factorization (SNMF) method. The SNMF function was defined as the following:

$$min_{W,F} \frac{1}{2} ||Z - WF||_{Fro}^2 + \lambda \sum_{t=1}^{T} ||F(:,t)||_1^2 \qquad s.t. W, F \geq 0 \quad (3)$$

Fro is the Frobenius norm, and $\lambda$ is the regularization parameter estimated with the trade-off between accuracy and sparse degree (as selected by cross-validation). The results of this study showed that in online evaluation, throughput for the SNMF method is significantly higher than simple NMF and linear regression. Also, the completion time for the proposed method is shorter than NMF and linear regression. In another study, Kim *et al.* [95] proposed using Hadamard product to improve the NMF. Moreover, Yang *et al.* used the combination of the Hadamard product and constrained NMF (CNMF-HP) to improve the NMF method [90]. The performance of the proposed method was investigated both offline and online. According to this study, the CNMF-HP method was better than NMF, SCNMF, and NMF-HP in the offline (in terms of average SNR and $R^2$ indices) and online (in terms of throughput indices) evaluation.

Although good accuracy has been obtained in controlled laboratory conditions, none of these methods have yet been used commercially. Therefore, in future studies, methods should be evaluated in real conditions instead of increasing the laboratory's accuracy [19].

### *3.4. Pattern recognition-based myoelectric control*

#### *3.4.1. Classification-based control*

In general, the process of pattern recognition is followed according to Figure.

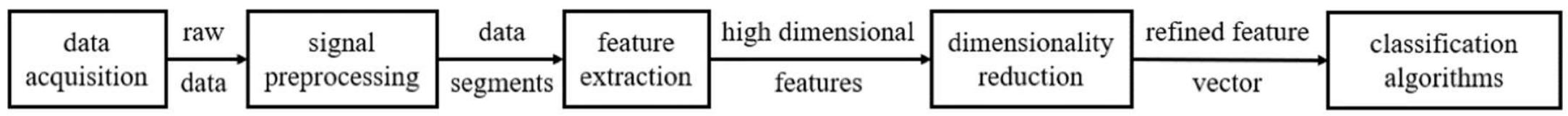

**Figure. 16.** EMG pattern recognition's path. (Reproduced with permission from [96])

After sEMG data acquisition, pre-processing (such as filtration, segmentation), feature extraction is performed for each signal segment (e.g., time, frequency, or time-frequency domain features) [24]. The time-domain features are usually preferred over frequency or time-frequency domain features because they do not require high processing time and high dimensional vectors [88]. These features help to perform pattern recognition and movement classification. The output of classifiers enters the controller as input and moves the prosthesis to the intended position. Sometimes, the feature selection or reduction step in the pipeline decreases the computational complexity and increases performance accuracy simultaneously [97].

In 2003, Soares *et al.* [98] used an autoregressive (AR) model and Artificial Neural Network (ANN) classifier for sEMG classification and virtual upper-limb prosthesis control. The Autoregressive model was used because the sEMG signal is inherently nonlinear, nonstationary, and stochastic [88]. The achieved accuracy was 100% and 95% with 10 and 4 order AR. Such models were robust against sEMG amplitude and phase variations.

In 2005, Huang *et al.* [99] presented a study in which 12 intact subjects participated in the experiment. Three feature sets, including TD Hudgin's features (MAV, WL, ZC, and SSC [45]), RMS + autoregressive (AR), and TD Hudgin's features + RMS + AR, were selected. Gaussian Mixture Models (GMMs) classifier was used, and classifying error for each feature set was 5.8%, 3.72%, and 3.09%, respectively. For comparison, LDA, MLP, and LP classifiers were applied on RMS + AR feature set, and the classification error for each classifier was 4.42%, 4.62%, and 4.53%, respectively.

In 2007, Farrell *et al.* [113] evaluated maximum acceptable response time (required time for EMG acquisition and analysis) for real-time myoelectric control systems. Among 20 healthy subjects, repeated measures ANOVA statistical analysis and a linear mixed-effects model showed that the optimal delay time is between 100 ms and 300 ms. Therefore, sEMG epochs length was significant. In 2003, Englehart *et al.* [114] similarly investigated the delay time and classification accuracy with different window lengths and reported that pattern recognition-based classifier systems' accuracy decreased when decreasing window length.

In 2008, Oskoei *et al.* [100] compared TD Hudgin's (MAV, WL, ZC, and SSC) and RMS + AR features with LDA, SVM, and MLP classifier and different segmentations. The SVM classifier with TD Hudgin's features showed stable performance in window length variation, fast response, and high-performance accuracy (95.5 ± 4%).

In 2013, Al-Timemy *et al.* [101] studied finger movements offline classification with multi-channel sEMG data recording from forearm muscles. Time Domain-Auto Regression (TD-AR) feature was extracted. The EMG channel was reduced to 6 channels (at feature reduction step) using Orthogonal Fuzzy Neighborhood Discriminant Analysis (OFNDA), and the LDA classifier was used for classification. Overall, the proposed system's accuracy for 15 finger movements was 98% in 10 healthy subjects. It was 90% for 12 finger movement classes in 6 amputees.

In 2011, Simon *et al.* [103] performed a study on real-time, closed-loop, and multiple-degree freedom (1- and 3-DOF) classifiers for myoelectric control systems. Five amputees participated in this study. The experiment was performed to move a virtual arm to the considered posture in 3 different conditions (1- or 3-DOF and with 1 or 3 required movements of 7 movement classes) and eight repetitions for training and testing phases. The sEMG signal was windowed, and the mean absolute value (MAV), the number of Zero Crossings, Waveform Length, and the number of Slope Sign Changes (SSC) features were extracted from each window. They achieved 97.2 ± 2.0% classification accuracy for 1DOF and a required movement condition and 94.1 ± 3.1% for 3DOF and 1 or 3 required movements using LDA classifier.

In 2012, Jiang *et al.* [115] performed a study on proportional and simultaneous prosthesis control using MLP classification. The sEMG signal was recorded from amputees and able-bodied subjects during wrist movements with 3DoF and time-domain autoregressive features extracted from overlapped epochs. The overall accuracy on healthy subjects was 72.0% ± 8.29% and on amputees was 62.5% ± 8.50%.

In 2013, Phinyomark *et al.* [24] compared 50 EMG time, and frequency domain (TD and FD) features for ten hand movements classification. A healthy subject participated in this study. sEMG signal was used in this study that was recorded from 4 electrodes. The sEMG data were recorded for 21 days to investigate the effect of long-term usage or reusability. The results showed that Sample Entropy (SampEn) with 93.37% accuracy was the best robust single feature compared with Hudgin's TD features. Also, SampEn, Cepstrum Coefficients (CC), RMS, WL were the best robust multiple-feature set with an accuracy of 98.87%. They also showed that LDA was better than other classifiers such as RFS, MLP-NN, SVM, and KNN in classifying fluctuating EMG signals. Moreover, the appropriate segmentation was 500 ms with 125 ms overlapped segmentation in real-time and high-performance systems.

In 2014, Ameri *et al.* [83] did a similar study on two amputees and ten healthy subjects based on an SVM classifier. They compared the performance of their system with the ANN. The throughput, completion rate, path efficiency, and overshoot parameters were analyzed. The SVM classifier showed better performance in healthy subjects and better path efficiency and throughput in amputee subjects. The SVM method had lower training and testing time and was closer to real-time.

In 2017, Beyda *et al.* [104] designed a hand prosthesis to imitate six finger movements by classifying registered sEMG of the forearm muscles. At first, the sEMG signal was rectified and smoothed (with 50-500Hz bandpass filter) and then was windowed and time-domain features of each window including signal Energy (E), Maximum value of the signal (Max), Variance of signal (VAR), Average Rectified Value (ARV), and RMS were extracted. In this study, the fuzzy logic algorithm was used as a classifier, and the sum of each window's features was the algorithm's input. The mean performance accuracy of this system was 90.87 ± 9.13%.

In 2017, Benatti *et al.* [105] designed a low-cost, real-time, and poliarticulated prosthetic hand. Five healthy and four amputee subjects participated in this study. They suggested hardware for sEMG signal acquisition with Ottobock 13E200 sensors for raw EMG signal time samples (Figure). The on-board Support Vector Machine (SVM) model was used to classify hand gestures. This method was also tested on the Ninapro dataset and obtained an accuracy of 91.83% (offline). The method's accuracy was 87.37% and 89.09% for healthy and amputee subjects in online tests, respectively.

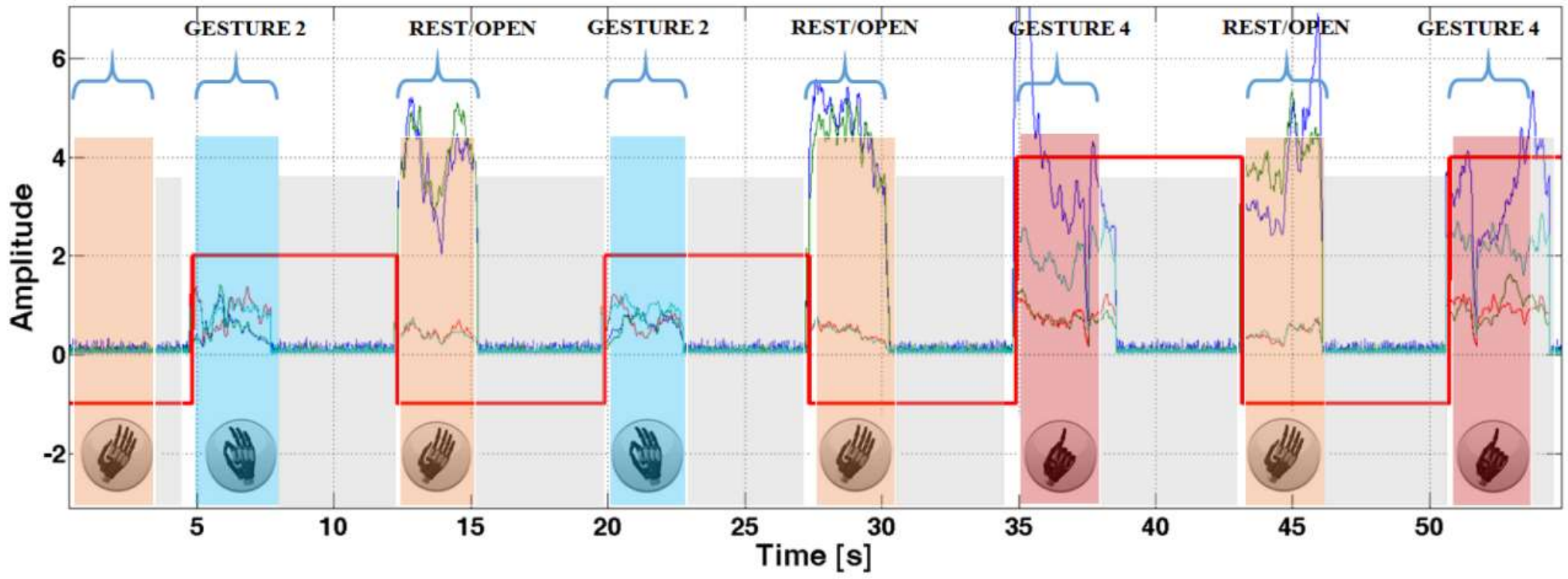

**Figure. 17.** sEMG signal during hand movements (Reproduced with permission from [105]).

In 2018, Rehman *et al.* [106] used EMG raw data (after filtering and overlap windowing) as input to a Convolutional Neural Network (CNN) classifier. Seven able-bodied subjects participated in this study to performed six hand movements (in addition to resting state), and long-term bipolar sEMG signals were recorded for 15 days with Myo armband. For comparison, they used Linear Discriminant Analysis (LDA), Stacked Sparse Autoencoders with features (SSAE-f), and raw samples (SSAE-r) classifiers with Mean Absolute Value (MAV), Waveform Length (WL), Slope Sign Change (SSC), and Zero Crossing (ZC) as inputs. A comparison was performed by two-way ANOVA statistical analysis. It was shown that the CNN and SSAE-f were significantly higher than LDA and SSAE-r methods. Also, CNN's training time was shorter than SSAE-f.

In 2018, Patel *et al.* [108] used the Cosine Similarity (CoS) classification method based on the regularity in Muscle Activation Patterns (MAPs vector in each sEMG segment = (MAV1, MAV2, …, MAVd) where *d* is the number of sEMG channel) and used muscle coordination for classification. They also compared their method with LDA during offline and online classification. The MAPs obtained by a vector of EMG amplitudes were highly correlated to MVC levels in each movement. The sEMG data were recorded from eight electrodes in seven healthy subjects' forearms in an offline experiment and 11 healthy and amputee subjects in an online experiment with right-handed Michelangelo prosthesis (Otto Bock, Vienna). The Experimental setup included four paired movements and one relaxed-hand state with two levels of MVC. They used MAV features for CoS, while MAV and full time-domain feature (TD) was used for the LDA classifier. The offline experiment's Cos classifier classification error was either lower or higher than LDA, depending on the chosen feature for the LDA classifier. However, in an online experiment, CoS showed better performance than the state-of-the-art LDA classifier in three healthy subjects' movements and amputees' entire movements. This method showed robust and reliable performance during functional tasks, especially at online experiments, and better classification performance for amputees' functional tasks while utilizing less training data, smaller feature sets, and lower training time than the LDA method.

In 2018, Ameri *et al.* [84] compared the CNN method's performance to SVM for real-time and simultaneous myoelectric control pattern recognition. The CNN classifier's average accuracy was 91.61 ± 0.39, and the SVM classifier was 90.63 ± 0.31 in offline 4-fold cross-validation. One-way ANOVA analysis showed that there was not a significant difference between the offline and online results. However, it is worth mentioning that the CNN method was applied to the raw sEMG data instead of TD and frequency features.

In 2018, Phinyomark *et al.* [112] extracted 26 features in the time and frequency domain for evaluating the effect of sampling rate on hand movement classification. They divided those features into eight groups, among which a set was the Hudgin's features (MAV, WL, ZC, and SSC). Four different datasets were obtained from 31 able-bodied subjects and nine transradial amputees. Results showed that decreasing the sampling rate decreases the accuracy.

In 2018, Ding *et al.* [109] proposed a parallel multiple-scale convolution architecture, a deep learning method, to decrease information loss during feature extraction and increase recognition accuracy. Their proposed method had a larger size of kernel filter than other CNN-based hand recognition methods. Seventeen hand movements of the DB2 Ninapro database were used, including six repetitions of 50 movements from 40 intact subjects recording by 12 sEMG electrodes [18]. No preprocessing procedures such as low-pass filtering, fast Fourier transform (FFT), and standardization for preventing information loss and time delay were used. Overlapped sEMG segments of 100 ms were used and were transferred into sEMG images (**Error! Reference source not found.**). The average accuracy of the method was 78.86% for the entire DB2 gestures. The results showed that the parallel multiple-scale convolution architecture with larger sizes of kernel filter (C-B1PB2) increased the classification accuracy (**Error! Reference source not found.**) compared to sEMG characteristics such as muscle independence [109].

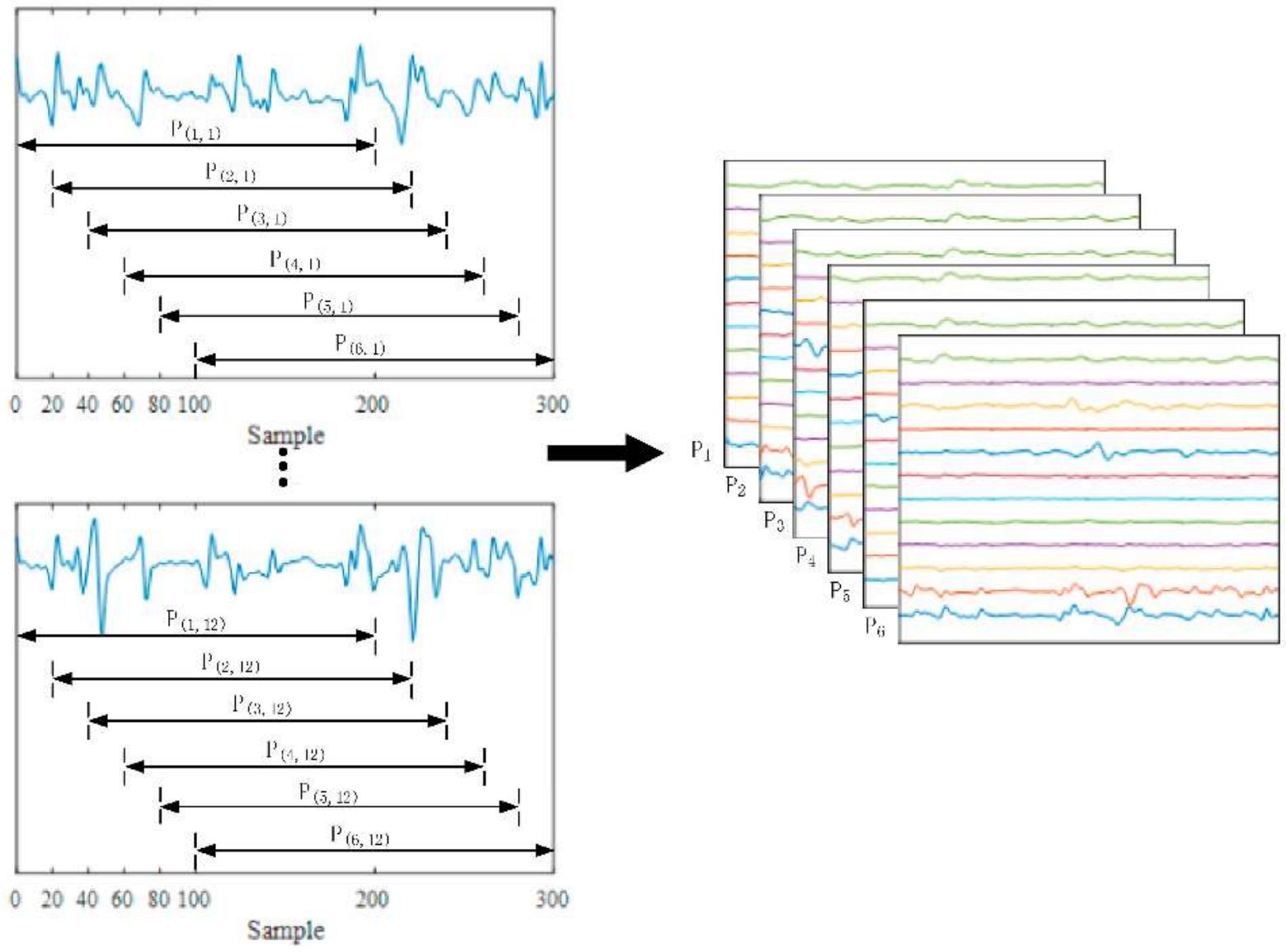

**Figure. 18.** Producing sEMG image from sEMG segments for classification system input. (Reproduced with permission from [109]).

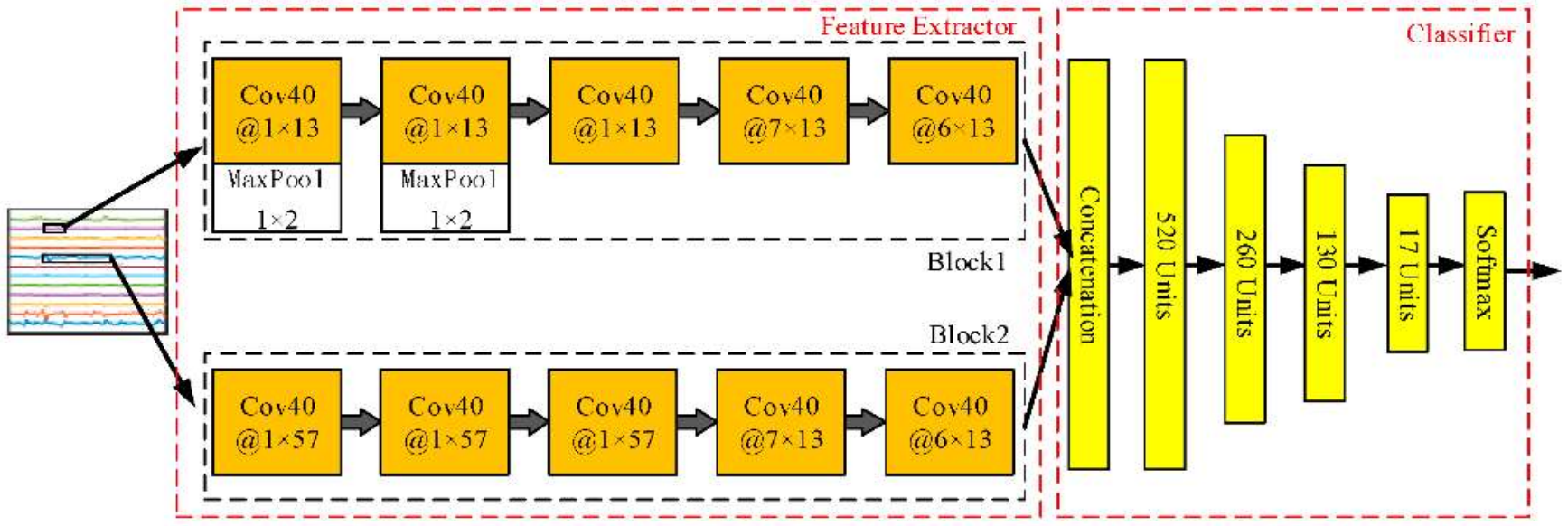

**Figure. 19. The** C-B1PB2 architecture for sEMG image classification. (Reproduced with permission from [109]).

In 2019, Zhang *et al.* [110] recorded sEMG data from 12 subjects for five hand movements using Myo armband. The sEMG signal was filtered by a 5-Hz high pass filter after rectification. Then, five time-domain features include MAV, Slope Sign Change (SSC), Waveform Length (WL), Root Mean Square (RMS), and Hjorth Parameter (HP), were extracted. The ANN classifier was applied to the features for the training and testing phase. The average recognition accuracy was 98.7%.

In 2020, Asif *et al.* [107] evaluated the CNN classifier's performance on ten active upper extremity movements of 18 healthy subjects. It is necessary to tune hyper-parameters in deep learning methods such as training iteration numbers and the training phase's learning rate. Also, as the learning rate decreases, the number of training iterations should increase to achieve higher accuracy and network convergence. An average classification error of 10.4% was observed with a learning rate of 0.001.

In 2020, Raheema *et al.* [102] extracted five time-domain features (Root Mean Square (RMS), Standard Deviation (SD), Maximum, and Minimum) and used Multi-Layer Perceptron (MLP) neural networks to classify five hand movements (Figure). In this study, The Myo armband device was used for sEMG recording. They obtained an average accuracy of 99% with the MLP classifier for five different classes of hand gestures.

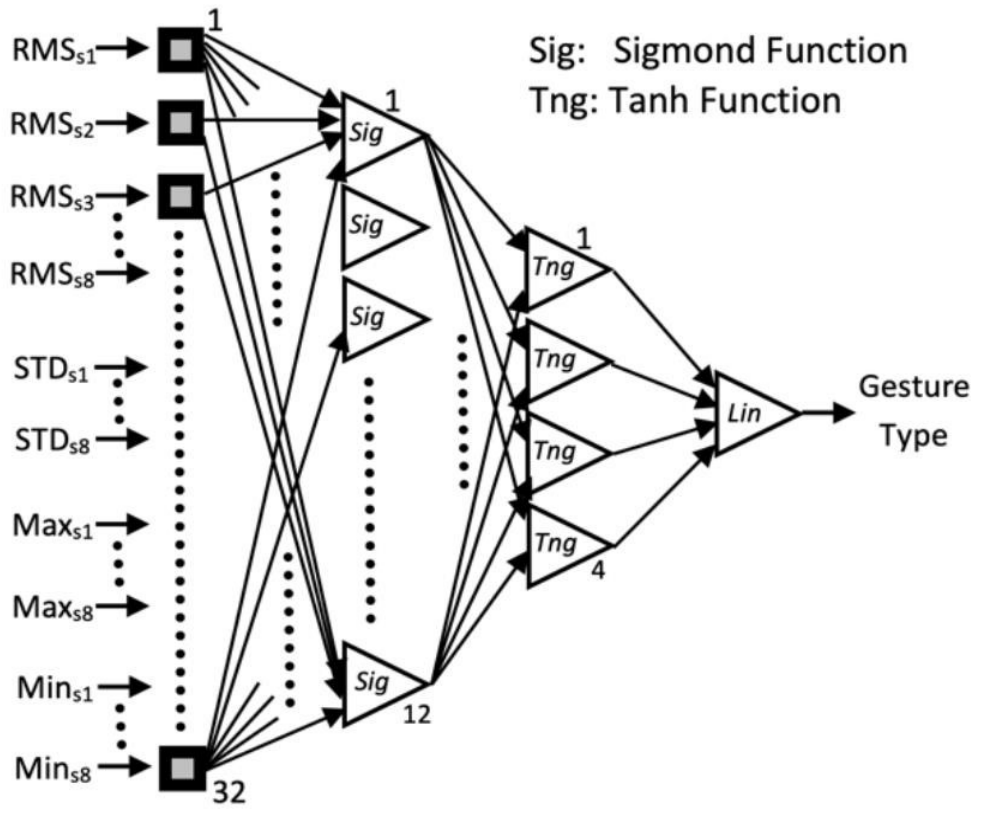

**Figure. 20.** The MLP neural network diagram for gesture classification. (Reproduced with permission from [102])

In 2020, Hassan *et al.* [111] used sEMG signals recorded by Myo armband from six healthy subjects to move a 5-DoF Aideepen ROT3U robotic arm real-time. They used a pattern recognition system including segmentation, feature extraction, and classification to analyzed and distinguished seven hand movements. After segmentation, six time-domain (TD) features (Mean Absolute Value (MAV), Waveform Length (WL), Root Mean Square (RMS), Autoregressive Coefficients (AR), Zero Crossings (ZC), Slope Sign Changes (SSC)) were extracted from each segment. Then, a comparison was made between Support Vector Machines (SVM), Linear Discriminant Analysis (LDA), and K-Nearest Neighbor (K-NN) classifiers. Results showed that the SVM classifier with 95.26% accuracy had higher performance accuracy than LDA (92.58% ) and K-NN (86.41% ). Also, it was shown that the window size has an essential effect on the system's accuracy.

*3.4.2. Regression-based control*

Regression is a technique for estimating continuous outputs from continuous input features. The main difference between regression and classification methods is that instead of deciding on a particular class, a continuous (proportional) output value is estimated for each degree of freedom (DoF). Such control allows natural and smooth control on prostheses [90].

Shehata *et al.* [116] compared the classifier-based and regression-based approaches as two commonly used myoelectric control strategies. The performance test results indicated that the filtered classification-based controller had better short-term performance in path efficiency. However, regression approaches enabled the development of a stronger internal model, which means better long-term performance. They suggested that the improvement of the myoelectric control system's human understanding is possible by rich feedback associated with regression control.

In 2014, Hahne *et al.* [86] compared four linear and nonlinear regression techniques, including linear regression (LR), the mixture of linear experts (ME), multilayer-perceptron, and kernel ridge regression (KRR) for independent proportional and simultaneous myoelectric control of wrist movements with two DoF. The results showed that KRR is superior to other methods. Such non-parametric models suffer from high processing costs and high memory requirements. However, by linearizing in the feature space, a simple regressor like ME could achieve similar performance and solve the computational cost of KRR. They also showed that recording training data for all possible combinations of DoFs is not necessary if the models are well-generalized to DoF-combinations.

Regression-based techniques allow the user to control multi DoF simultaneously. Hahne *et al.* [117] evaluated a two-DOF linear regression (LR) on subjects with transradial amputation or congenital limb deficiency to demonstrate the feasibility of robust simultaneous and proportional prosthetic hand control. They also compared the regression-based technique with two conventional control two-DOF control approaches, co-contraction control (CC) and slope control (SC). These techniques switch the active DOF to control more DOF. Their outcomes showed LR approach outperformed conventional control. They also observed that LR performance does not degrade when the regression model was trained on the first day and tested on the second day.

Controlling position and velocity of movement are two strategies used for proportional myographic control. Velocity-control needs less overall effort for a user and no limitations of the range of motion. In position control, the effort to contract the muscle is mapped to the device position or angular displacement, so the user should maintain muscle contraction to retain the position. It can cause muscle fatigue. To overcome this problem, prosthetic devices use velocity instead [83,118]. In 2019, Igual *et al*. [118] presented a novel general control concept that is not limited to either position or velocity control. They used the current position to predict the next intended location, e.g., an auto-regressive predictor. They used closed-loop, real-time feedback to adapt humans and machines simultaneously.

Amsuess *et al.* [119] proposed a technique that can automatically switch between sequential (one DoF at a time) or simultaneous (multiple DoF) prosthesis control. They used regression ability for fast and natural coarse positioning of the hand and classification for precise single-DOF control. Their method could decide whether the user intended a combined-DOF or single-DOF movement to seamlessly switching between two expert methods for estimating movement situation (regression for two DoF and classification for one DoF).

Ameri *et al.* [120] validated the usability of regression Convolutional Neural Network (CNN) in simultaneous wrist motion estimation from online EMG recording. They compared their method with the support vector regression (SVR)-based method [83]. They showed that the CNN model could better extract underlying motor control information from EMG signals during single and multiple DoF tasks due to higher regression accuracies, especially with the higher EMG amplitudes.

To evaluate linear regression-controlled prostheses' effectiveness in daily life, Hahn *et al*. [121] conducted a study on a subject who had a transradial amputation 35 years ago. They used an Otto Bock Michelangelo hand for about 12 months. This prosthesis was used as a baseline with conventional two-channel control. They evaluated the system with functional tests and a questionnaire at the beginning and the end of the experiment. For this participant, the regression-controlled prosthesis outperformed the conventionally controlled method. Finally, the study showed that the control of prosthesis based on linear regression is not limited to laboratory studies and can be used daily.

Using prostheses in life outside the laboratory may cause non-stationary conditions. Factors such as changes in the electrode-skin interface due to sweating, changes in arm-position, and time between the training and testing can adversely affect myo-control performance over time. In 2017, Hahn *et al.* [122] investigated real-time comparison between classification-based (linear discriminant analysis (LDA)) and regression-based (linear regression (LR)) myo-control in non-stationary conditions by adding white noise to one of the EMG channels. The results showed that in the offline experiment, the addition of noise was similarly affected in both methods; however, with the user in the loop (online), the regression-based approach was significantly less influenced by changes in signal characteristics than the classification approach. This result showed the superiority of recently proposed regression schemes for myoelectric control over more established classification approaches in non-stationary conditions.

Strazzulla *et al.* [123] proposed a nonlinear machine learning technique to predict each motor's expected torque of two robotic hands. The learning model was based on an incremental variant of the Ridge Regression algorithm (iRR) coupled with Random Fourier Features (RFF) to simultaneous and proportionate control of ten independent DoFs (five for single-hand and five for bimanual). According to their results, only five minutes of retraining were needed for a total session duration of about one hour and forty minutes. They accomplished a completion rate of 95% of single-handed tasks and 84% of bimanual tasks.

Kapelner *et al.* [124] compared linear regression performance on traditional time-domain sEMG and neural features (motor unit discharge timings) obtained by sEMG decomposition. The results showed that neural features outperformed traditional sEMG features. They also demonstrated that movement speed and regression performance based on neural features was pertaining to the subject. Table 1 summarized some recent studies on regression-based myo-control.

**Table** 1. Recent studies of regression-based myoelectric control

| ref | Goal of study | Regression mythology | DoF | Evaluation population | Extracted feature | EMG system |
|---|---|---|---|---|---|---|
| **(Jiang et al. 2014) [125]** | Investigating the impact of mapping of EMG Signals on Kinematics for a precise myoelectric control | Linear regression | 2 | Nine able-bodied | Envelopes of the surface EMG | commercial biosignal amplifier (EMGUSB2, OT Bioelettronica, Italy) with Sixteen monopolar surface electrodes (Neuroline 720, Ambu, Denmark) |
| **(Ameri et al. 2014) [83]** | comparison of support vector regression and artificial neural network | Support vector regression | 3 | 10 able-bodied + 2 transradial limb deficiency | Mean absolute value<br>Waveform length<br>Zero crossings<br>Slope signs changes | Eight<br>bipolar wireless surface electrodes (Delsys Inc.) |
| **(Hahne et al. 2014) [86]** | comparison of linear and nonlinear regression techniques | Linear regression<br>Mixture of linear experts<br>Multilayer-perceptron<br>Kernel ridge regression | 2 | 10 able-bodied + 1 congenital upper<br>limb deficiency | Variance<br>Log-variance<br>Root-mean-square | biosignal amplifier (OT Bioelettronica<br>EMGUSB-2) with 192-channel electrode grid (ELSCH064NM 3–3, OT Bioelettronica, |
| **(Hahne et al. 2015) [126]** | exploring co-adaptive real-time learning as a tool to train regression algorithms for myoelectric control | linear regression | 2 | 10 able-bodied+ 2 congenital limb-deficiency | Log-variance | Dry electrodes (g.tec USBamp g.tec Sahara) |
| **(Smith et al.2015a) [127]** | Decoding patterns of muscle co-activation and evaluating the ability of linear regression | Linear regression | 3 | Eight able-bodied | Mean absolute value | Fine-wire EMG electrodes (Motion Lab Systems, Inc.) |
| **(Smith et al. 2015) [128]** | Evaluating probability weighted regression system | Probability-weighted linear regression and Linear regression | 3 | Eight able-bodied+<br>Two transradial amputee | Mean absolute value<br>Waveform length<br>Slope-sign changes<br>Zero crossings<br>model parameters | Fine-wire EMG electrodes (Motion Lab Systems, Inc.) |
| **(Hwang et al. 2017) [129]** | Effect of arm position change and donning/doffing on controllability | Linear regression | 2 | 14 able-bodied + One congenitally deficient upper limb | Root-mean-square | 16 channel biosignal amplifier (g.USBamp, g.tec Inc., Graz, Austria) with a dry active electrode system (g.SAHARA) |
| **(Strazzulla et al. 2017) [123]** | predicting the expected torque of each motor of two robotic hands by proposing a nonlinear machine learning technique | Ridge Regression algorithm with Random Fourier Features | 10 | Nine able-bodied | Filtered signal row | sEMG electrodes (model MyoBock 13E200, Otto Bock, Duderstadt, Germany) |
| **(Hahn et al. 2017) [122]** | Investigating robustness and performance in a closed-loop application of linear discriminant analysis and linear regression | Linear regression | 2 | Ten able-bodied + One transradial amputee | Root mean square | Eight dry, active Otto Bock EMG electrodes (13E200AC) |
| **(yang et al. 2018) [130]** | Providing wrist motion/force synchronously platform | Support vector regression | 3 | Six able-bodied | Root mean square | eight wireless surface EMG electrodes (Trigno, Delsys INC, USA) |
| **(Hahn et al. 2018) [117]** | Evaluation of linear regression under challenging conditions | Linear regression | 2 | Five transradial amputee or congenital limb deficiency | conditioned EMG envelopes | Eight conventional electrodes (13E200, Otto Bock) |
| **(Shehata et al.2018) [116]** | comparison the classifier-based and regression-based technique | Linear Regression<br>Support Vector Regression | 2 | 24 able-bodied | Time-domain feature | UNB Smart Electrode System |
| **(Ameri et al. 2019) [120]** | validation the usability of regression Convolutional Neural Network in simultaneous wrist motion estimation based on online EMG | Regression Convolutional Neural Network<br>Support vector regression | 2 | Ten able-bodied | Signal row for Convolutional Neural Network<br>Mean-absolute value<br>Waveform length<br>Zero-crossings<br>Slop sign changes<br>Mean frequency | eight pairs of surface electrodes (g.HiAmp, g-tec Inc.) |
| **(Kapelner et al. 2019) [124]** | Predicting joint angle of the wrist from motor unit discharge timings | Linear regression | 3 | Five able-bodied+<br>One transradial amputee | Root mean square<br>Slope sign changes<br>Zero crossings<br>Waveform length | EMG amplifier (EMGUSB2, OT Bioelettronica) with high-density electrode grids (ELSCH064NM3, OT Bioelettronica) |
| **(Dewald et al. 2019) [131]** | Evaluating the potential effectiveness of chronically implanted intramuscular electrodes for enhancing multi-DOF prosthetic hand functionality | K Nearest Neighbor regression | 3 | One transradial amputee | Mean absolute value<br>Waveform length | Ripple Grapevine Neural Interface Processor (NIP) system with an EMG front end for the intramuscular Myoelectric setup and a |
| | | | | | | Touch proof adaptor for the surface EMG setup (Ripple, Salt Lake City, UT) |
| **(Igual et al. 2019) [118]** | Presenting combinations of position and velocity control to emerge naturally from a closed-loop interaction of the human and machine | Linear regression | 2 | 15 able-bodied + 1 amputee+ 1 congenital | Log-variance | Myo Armband from Thalmic |
| **(Hahn et al. 2020) [121]** | evaluation the effectiveness of linear regression-controlled prostheses in daily life | Linear regression | 2 | One transradial amputee | Signal row | Eight Otto Bock 13E200 electrode |

### *3.5. Deep learning for prosthetic control*

Some deep learning methods were discussed in the pattern-recognition methods. In this section, we focus on such methods in detail. Artificial Intelligence (AI) is used where a computer imitates cognitive processes, such as learning and problem-solving. On the other hand, machine learning (ML), a concept defined by Samuel in 1959 [132], is an AI subfield. Deep learning (DL) has arisen among the most promising techniques under the ML umbrella [133]. Figure 21 represents a brief overview of AI development.

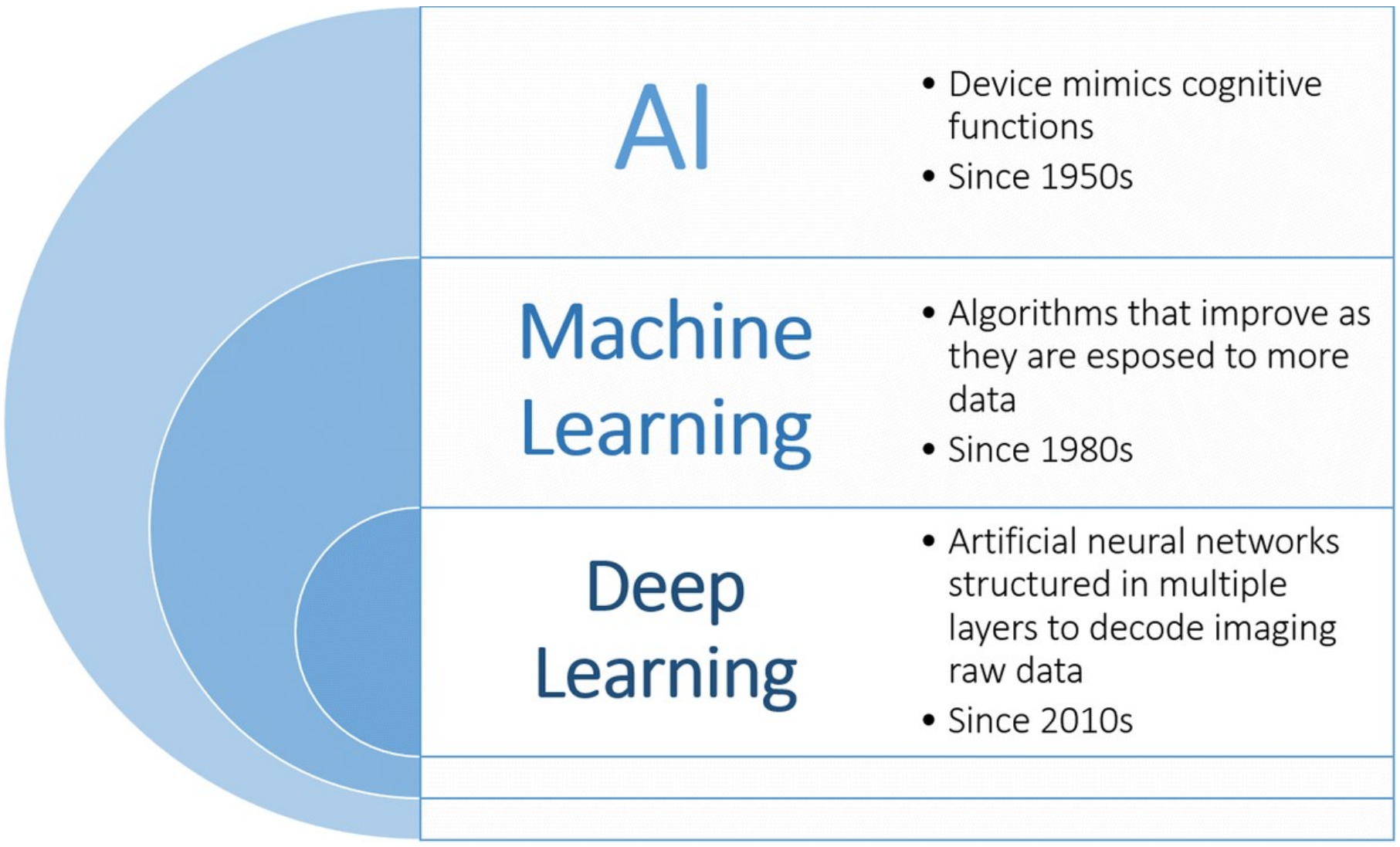

**Figure 21.** A brief history and relation between AI, ML, and DL. (Reproduced with permission from [134])

Owing to the growing release of publicly available benchmark EMG databases [135–138], researchers have begun to explore the power of DL to process sEMG data. Nevertheless, having a robust framework for gesture estimation is challenging due to the electrode positioning, wearing wires and switches [139], and signal alterations linked to inactivity-induced atrophy [140].

Transfer learning (TL) is a popular method in deep learning, where information of a pre-trained network is exchanged or passed to other networks for the extraction and classification of features. It can also handle imbalance or limited data and extract useful features in less time than conventional DL approaches. Ameri *et al.* [25] applied a customized transfer learning method based on Convolutional Neural Network (CNN), which is pretrained on no-electrode-shift data for predicting flexion-extension and pronation-supination after 2.5 cm electrode displacement. They utilized ten channels and obtained 21.5 ± 2.3 % and 46.0 ± 4.1 % error rate after electrode shift for flexion-extension and pronation-supination. Demir *et al.* [141] acquired 99.04% accuracy with AlexNet fc6 + AlexNet fc7 + VGG16 fc6 + VGG16 fc7 deep feature concatenation and SVM classification for classifying ten different physical actions on the UCI dataset [142]. Figure 22 shows an illustration of their proposed method.

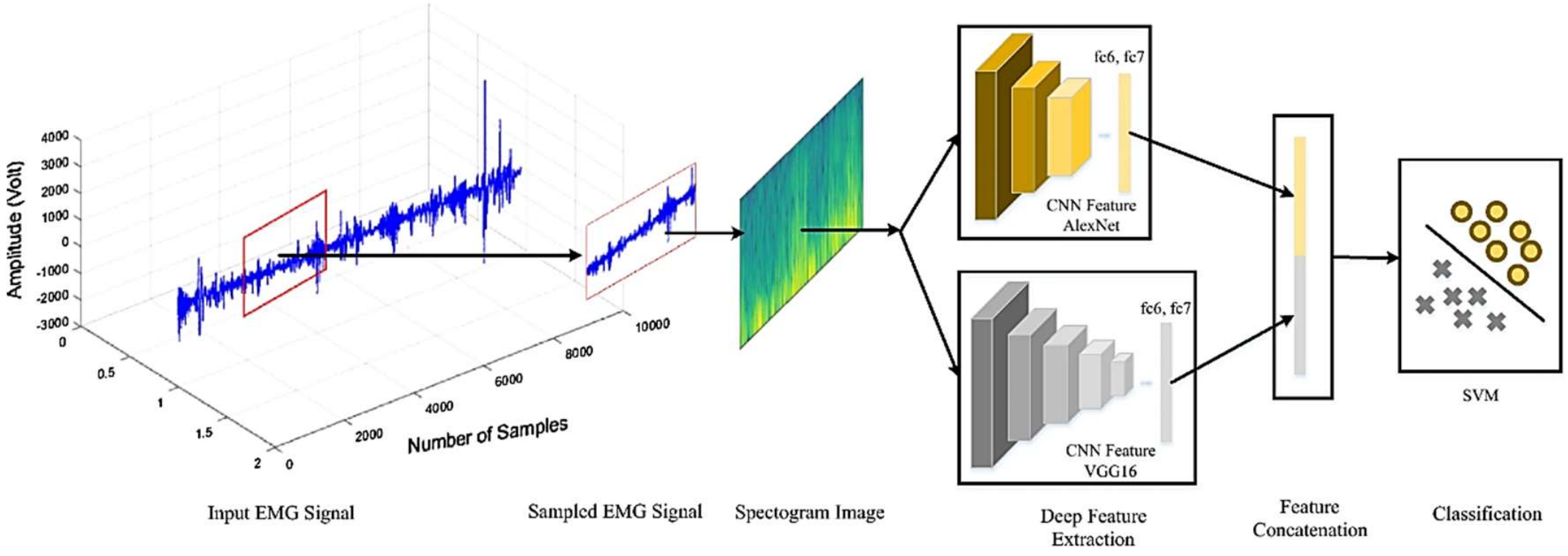

**Figure 1.** Deep feature extraction using transfer learning and SVM classification methodology for EMG signal classification. (Reproduced with permission from [141]).

Using CNN, Atzori *et al.*[143] classified 50 hand gestures on 67 intact subjects and 11 amputees. Three NinaPro datasets were used, and 66.59 ± 6.4%, 60.27 ± 7.7%, and 38.09 ± 14.29% accuracy were achieved. Their findings indicate that CNN with a straightforward architecture was equivalent to the average classical machine learning classification methods. Park and Lee [144] also used NinaPro sEMG dataset to decode six hand movements and achieve 60% accuracy using CNN. DL's object detection is utilized for grasp classification by Ghazaei *et al.* used [145] and achieved 84% accuracy. Instead of applying raw sEMG data on CNN, Ulysse *et al.* [146] used continuous wavelet transforms of sEMG as the input of the CNN model for gesture detection and improved gesture detection performance from 68.98% to 98.31%. Triwiyanto et al. optimized CNN's hyperparameters, such as the number of convolution layer filters, dropout, type of optimizer, and the number of training data, and could achieve an average accuracy of 77% and 93% for 10-class gesture classification in different motion ranges [147]. A CNN-based self-calibration classifier that calibrated automatically over time was proposed by Zhai *et al.* [148]. For 10-movement experiments on intact subjects, they demonstrated 88.42 % accuracy using CNN compared to 87.86 % accuracy for support vector machines, although accuracy was 73.31 % and 72.01 % for amputees.

Hu et al. [149] proposed an RNN model for gesture detection by addressing the signal variability over time. It improved the accuracy of detection from 83.5% to 84.8% on the first sub-database of NinaPro (known as NinaProDB1), from 73.4% to 74.8% on the second sub-database of NinaPro (known as NinaProDB2), from 83.9% to 92.5% on a sub-database of BioPatRec (known as BioPatRec26MOV), from 97.7% to 99.7% on a sub-database of CapgMyo (known as CapgMyo-DBa) and from 92.1% to 94.9% on the CSL-HdEMG compared to the CNN framework that only modeled spatial information. The NinaPro dataset was also utilized by Wang *et al.* [150] to predict 20 joint angles using LSTM and achieve root mean square 5.98 ± 0.95 degrees. They also showed that LSTM outperformed radial basis function, a sparse Gaussian process using pseudo-input. Nguyen *et al.* [151] developed a new implantable platform for recording EMG and applied an RNN model on recorded data for detecting 19 gestures. Their proposed method comprises 21 convolutional layers, two recurrent layers (LSTM), two attention layers, and an output layer (Figure 23).

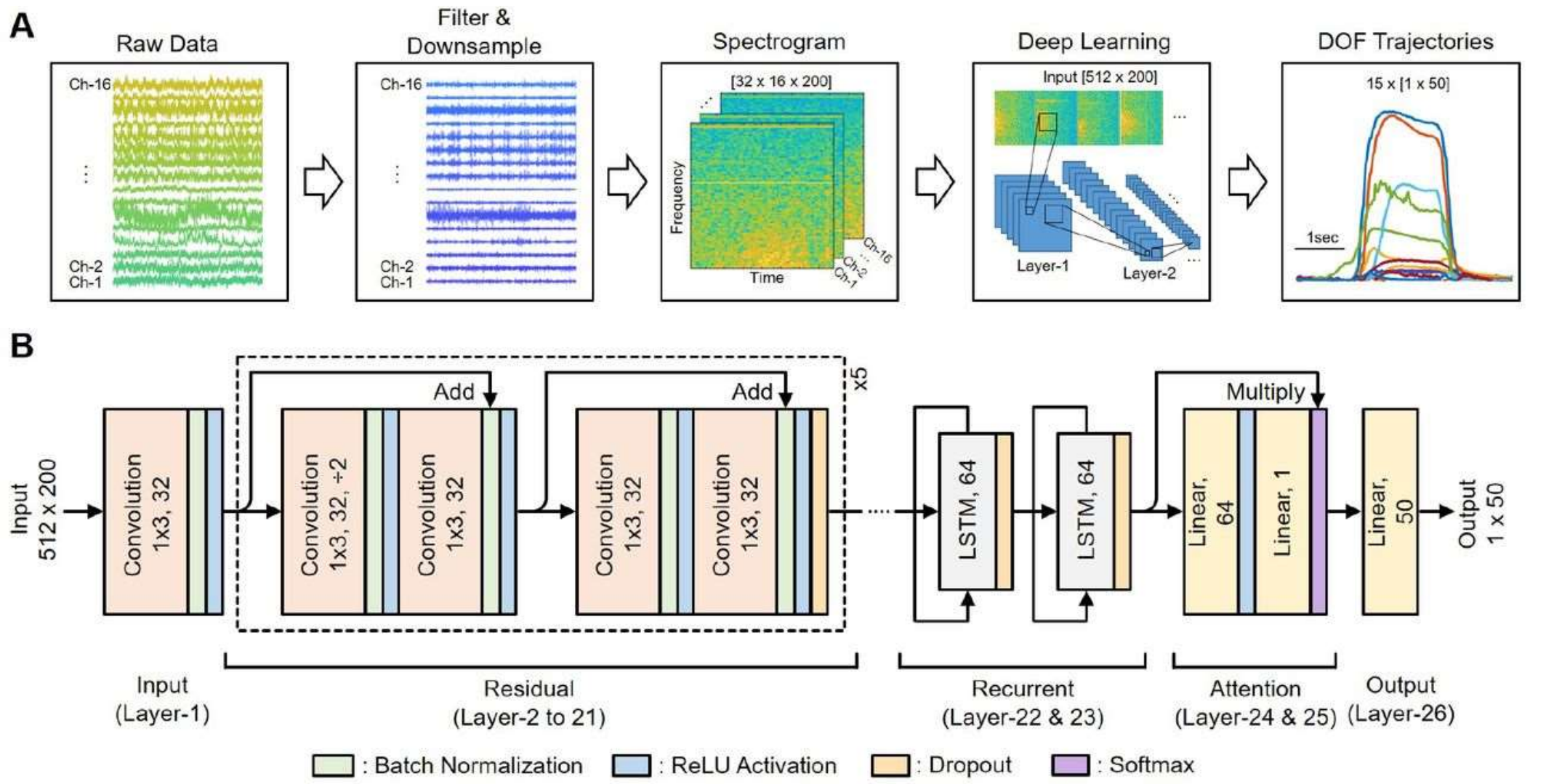

**Figure 23.** (A) Diagrams of the signal pre-processing flow, including filtering, feature extraction via spectrogram, and motor decoding using deep learning-based AI models. (B) The AI model design based on a 26-layer recurrent neural network (RNN) architecture includes residual, recurrent, and attention blocks. A total of 15 AI models with the same structure are trained for every DOF. Each model takes the input from all 16 recording channels. (Reproduced with permission from [151]).

Evidence of auto-encoder-based architectures is the study undertaken by Li *et al.* [152], which used PCA for feature reduction and used a two-layer stacked auto-encoder softmax classifier as the output layer. It was able to achieve more than 95 % accuracy on 15 reported sEMG for gesture detection. Also, Lv *et al*. [153] used a similar architecture for robust prosthetic control against electrode displacement. Using 192 HDsEMG channels, they obtained the error rate of 1.48 ± 1.43 % without electrode shift and 9.83 ± 8.15 % with 1 cm electrode shift. Jia *et al.* [154] proposed a deep learning algorithm that combined an auto-encoder for feature selection and a CNN for classification to distinguish ten groups of hand gestures. They achieved 98.1% accuracy, which outperformed other classifiers when no auto-encoder was used. Figure 24 showed the structure of auto-encoder and CNN that were used in their study.

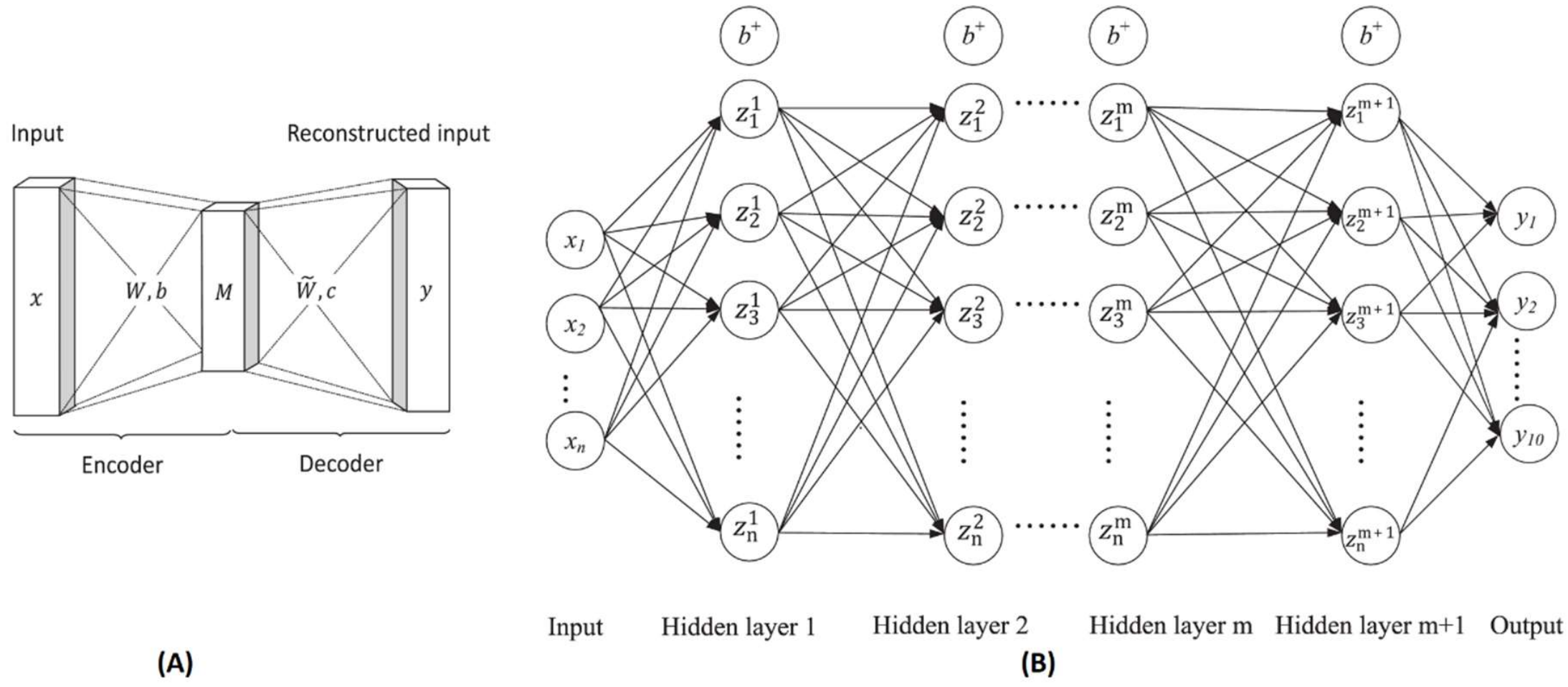

**Figure 24.** The structure of auto-encoder (A) and CNN (B) were used for classifying ten different hand gestures. The latent vector of the auto-encoder, the middle layer, was used as input to the CNN. (Reproduced with permission from [154]).

None of the deep learning and machine learning methods were found to be 100% effective. Misclassifications need to be remedied to make AI-enabled myoelectric hand control a valid option for an amputee. Otherwise, patients will get discouraged if they are ineffective in performing a mission, such as unintentional prosthesis gestures.

### *3.6. Performance indices*

There are different indices (metrics) used to describe various aspects of myoelectric control performance in practice. We provided a review of some of those indices with their definitions and their application in such studies.

In 2018, Farina *et al.* [155] experimented with using the autoencoding (AEN) method for online mapping of EMG signal into kinematics. Seven subjects with no neuromuscular disorders performed some myoelectric control tasks following the cursor on the user's screen in two degrees of freedom (DoF1: wrist flexion/extension and DoF2: wrist adduction/abduction) using the state-of-art (SOA) and AEN control algorithms. The performance of the method was assessed as the following:

- Completion time (or motion-completion time [125]) ($CT$, [s])

  Completion time is defined as the time spent by the subject to complete each task successfully.

- Completion rate (or motion-completion rate or success rate [156]) ([%])

  Completion rate is the number of completed tasks over all of the intended tasks.

- Overshoots ($k$)

  Overshoots is defined as the number of incidents where the arrow's tip is passed over the target before dwelling.

- Task index of difficulty ($ID$, [bit])

  Task difficulty is defined as the following using Shannon's extension of the Fit's law:

$$ID = \log_2\left(\frac{A}{W} + 1\right) \quad (4)$$

where $W$ and $A$ are the radius and the amplitude of the target, respectively. In this study, $W$ is fixed for the entire targets, and $A$ is defined as the following:

$$A = \left(0.5_{\gamma_1} + 0.5_{\gamma_2}\right)^2 \quad (5)$$

where $\gamma_1$ and $\gamma_2$ are the required angles to reach in DoF1 and DoF2, respectively.

- Throughput ($TP$, [bit/s])

  Throughput is the average ratio between the task index of difficulty and the completion time for each task defined as below:

$$TP = \frac{ID}{CT} \quad (6)$$

- Speed ($v$, [dp/s])

  Speed is defined as the ratio between the center of the cursor's trajectory length and the completion time of the task.

- Path efficiency ([%])

  Path efficiency is the optimal path ratio between the initial and the target points and the actual trajectory length.

Similar metrics were reported by Jiang *et al.* in 2014 [156] to analyze the online mapping between EMG signal and kinematics using three myoelectric control algorithms: linear regression (LR), nonnegative matrix factorization (NMF), and artificial neural networks (ANN). Nine subjects with no neurological disorders performed some wrist movements in two degrees of freedom (DoF) in which an arrow indicated the wrist flexion/extension (DoF1) by left-right movements and wrist rotation (DoF2) by rotational movements. Multi-channel surface electromyogram signals were recorded from the forearm during the tasks, and six online performance metrics were used to evaluate performance:

completion time ($t_C$), completion rate ($\alpha$), overshoots ($k$), speed ($v$), throughput ($TP$) and efficiency coefficient (Ã, [%]), which is defined similar to the path efficiency in [155].

Xiloyannis *et al.* [157] proposed a novel Gaussian process autoregressive approach in which both previous hand movements and instantaneous muscle activity were used to reach the prediction of future hand movements. Six healthy subjects were asked to perform some attempted tasks while the electromyography and mechanomyography of five muscles of the forearms were recorded, and 11 hand joints were monitored using a glove equipped with sensors. One of the critical indices that were reported was the lag.

- Lag ($\tau$)

  Lag is the time that the system demands to look in the past to predict the future. This metric affects the accuracy and the algorithm computations.

In 2014, Lobo-Part *et al.* [158] evaluated three control interfaces (EMG, force, and joystick) for active movement devices. Eight healthy subjects were asked to observe the moving target ($w$) and the cursor ($x$) in the computer monitor and try to minimize the distance between them ($e = w - x$) using the human-generated control signal ($u$) and one of the interfaces. In this closed-loop system ($H_{wx}(s)$), the human-interface system ($H_{eu}(s)$) gets the error between the input and the visual feedback and gives the control signal ($u$) to three control interfaces system ($H_{ux}(s)$). The control interfaces are mapped to the position of the cursor during the time. Fast Fourier transform (FFT) of x(t), w(t), u(t) and e(t) are X(f), W(f), U(f) and E(f), respectively. Thus, the power spectral estimated are defined as the following:

$$\hat{S}_{xx}(f) = X(f)X^*(f) \quad (7)$$

$$\hat{S}_{ww}(f) = W(f)W^*(f) \quad (8)$$

$$\hat{S}_{wx}(f) = X(f)W^*(f) \quad (9)$$

where hat and asterisk show the estimated value and the complex conjugate value, respectively.

- Tracking error ($\hat{F}_{ee}$)

  Tracking error is defined as the surface under the error signal's power spectrum (EQ. 10).

$$\hat{F}_{ee} = \sum_{i=1}^{n} \hat{S}_{ee}(f_i)\, \Delta\omega \qquad (10)$$

where $\hat{S}_{ee}(f_i) = \hat{S}_{ww}(f_i) - \hat{S}_{xx}(f_i)$ and $n = \frac{f_{max}}{NT}$ (11)

$N$, $T$, and $\Delta\omega$ are the number of samples, sampling time and frequency resolution, respectively and $f_{max}$ is the maximum frequency for which the tracking error is computed. Lower values of $\hat{F}_{ee}$ show more similarity of the power spectrums of cursor and target signals.

- Information transmission rate

  Information transmission rate quantifies the amount of output signal information of a communication channel related to the input signal (EQ. 12).

$$I = \int \log_2\left(1 + \frac{S(f)}{N(f)}\right) \mathrm{d}f \qquad (12)$$

where $S(f)$ and $N(f)$ are the signal and noise frequency contents, respectively. In this study, the human-interface system is considered as the communication channel, and the estimated information transmission rate is computed by EQ (13).

$$\hat{I} = \frac{1}{NT}\sum_{k} log_2\left(\frac{\hat{S}_{xx}(f_k)}{\hat{S}_{xx}(f_k) - \left|\hat{H}_{wx}(f_k)\right|^2 \hat{S}_{ww}(f_k)}\right) \qquad (13)$$

where $\frac{\hat{S}_{xx}(f_k)}{\hat{S}_{xx}(f_k) - |\hat{H}_{wx}(f_k)|^2 \hat{S}_{ww}(f_k)} = 1 + \frac{S(f)}{N(f)}$ and $\hat{H}_{wx}(f_k) = \frac{\hat{S}_{wx}(f_k)}{\hat{S}_{ww}(f_k)}$

- Estimated gain and delay parameters

The human-interface system modeled with the McRuer Crossover Model is described as the following:

$$H_{mod}(s,p) = k\mathrm{e}^{-\tau s} \quad (14)$$

where $s$ is the Laplace transform variable, the parameter vector $p$ is $p = [k, \tau]$, $k$ and $\tau$ are the gain and delay parameters, respectively. The following error cost function compares the frequency response functions (FRFs) of $H_{mod}$ and $H_{eu}$ to find the gain and delay parameters of the model to have a minimum error.

$$E(p) = \sum_k \hat{\gamma}_{wx}^2(f_k) \left| \ln\left(\frac{\hat{H}_{eu}(f_k)}{\hat{H}_{mod}(f_k)}\right)\right|^2 \quad (15)$$

where $\hat{H}_{eu}(f_k) = \frac{\hat{S}_{wu}(f_k)}{\hat{S}_{eu}(f_k)}$ and $\hat{\gamma}_{wx}^2(f_k) = \frac{|\hat{S}_{wx}(f_k)|^2}{\hat{S}_{ww}(f_k)\hat{S}_{xx}(f_k)}$

- Variance accounted for (VAF)

The variance accounted for shows the fitness of the model with the human-interface system. This time-domain metric is calculated using the mean estimated parameters of the human-interface system (EQ. 16).

$$VAF = \left(1 - \frac{var(\hat{y} - y)}{var(y)}\right) \times 100 \quad (16)$$

where, $\hat{Y}$ is the estimated output, and $Y$ is the measured output.

- Gain margin crossover frequency

The maximum frequency of tracking the target by the subject is called gain margin crossover frequency. This index is usually used to evaluate the stability margin in control engineering of closed-loop systems and is defined as the first frequency which gives a phase of $-180$ degrees to $\hat{H}_{wx}$.

In 2020, Kontson *et al.* [159] designed an experiment to analyze motor learning kinematic in the upper limb's body-powered bypass prosthesis. Six subjects with no disability of upper limb were asked to participate in the experiment and performed two trials of some tasks using the body-powered prosthesis to analyze movements of hand's joints in different degrees of freedom (DOF). Hand motions were captured with cameras and segmented into two sessions, and the following performance indices were defined:

- Resultant trajectories (T)

Resultant trajectories show the radial distance of each joint from the position whose entire joint angles were zero. This single vector was computed as the square root of the square of the vector sum of angular value ($\theta_x$) for each DOF ($x$) in a particular time (EQ. 17).

$$T(i) = \sqrt{\theta(\mathrm{i})_1{}^2 + \theta(\mathrm{i})_2{}^2 + \theta(\mathrm{i})_3{}^2} \quad (17)$$

- Normalized jerk (NJ)

Jerk is the derivative of acceleration, or in other words, the third derivative of position ($'''$), and normalized jerk which evaluates the smoothness is the time integral of jerk's square that is multiplied by $duration^5/Length^2$ to remove the effect of time and distance (EQ. 18).

$$NJ = \sqrt{\frac{1}{2}\sum_{i=1}^{N} T(i)'''^2 * \frac{duration^5}{Length^2}} \quad (18)$$

where $N$ is the length of the vector $T$.

- Path length (PL)

Path length indicates the distance tracked in three dimensions of joint angular trajectory for each segment (EQ. 19). Efficiency is shown by comparing this metric between different segments.

$$PL = \int_{i=1}^{N} |T(\mathrm{i}+1) - T(\mathrm{i})|\,\mathrm{di} \quad (19)$$

Maheu *et al.* [160] designed an experiment to evaluate the JACO robotic arm in 2011. Thirty-four participants with upper limb disabilities who used a wheelchair and no experience using robotic arms participated in the experiment. They were asked to perform a series of tasks with the JACO's joystick fixed on the wheelchair armrest. Thirty-one subjects could perform the entire task thoroughly. The participants' profiles were characterized to show which subjects and with which disabilities did the tasks successfully.

In 2006, Cavallaro *et al.* [161] developed the Myo-processor of the exoskeleton human-machine interface (HMI) using the Hill phenomenological muscle model and genetic algorithms for optimizing the parameters. The exoskeleton HMI predicted the torque of two upper limb joints (elbow and wrist) as the function of joint kinematic and the levels of neural activation. Three following metrics were reported in this experiment:

- Root-mean-square error ($E_{rms}$)

  The root-mean-square error of the reference torque $\tau(\mathrm{i})$ and the computed torque by the model $\tilde{\tau}(i)$ was defined as below:

$$E_{rms} = \sqrt{\frac{1}{N}\sum_{i=1}^{N}\left(\tau(\mathrm{i}) - \tilde{\tau}(i)\right)^2} \qquad (20)$$

  where $N$ is the number of samples.

- Percentage of time ($\eta_s$)

  The following equation was computed as the percentage of time where the error was less than a particular value.

$$\eta_s = \frac{\sum_{k=1}^{M} 1}{N} \qquad (\forall k |\ |\tau[k]| \leq s) \quad (21)$$

- Task execution time (TET)

  Task execution time was defined as the time took by the processor to complete a task.

In 2012, Muceli and Farina [48] proposed a method of proportional and simultaneous control of multiple DoFs using the ANN model to estimate kinematic of the hand from high-density surface EMG signals. EMG signal was recorded from six able-limbed subjects during the performing of mirrored bilateral movements. The mean relative error was used to indicate the performance of the proposed method.

- Relative error (rE)

  The relative error is the ratio between the measured and the estimated value and the range of movement.

In 2015, Huang *et al.* [162] experimented with introducing a cyber expert system (CES) for auto-tuning of the impedance control parameters of powered lower limb prosthesis compared to tuning by human experts (HMEs). Two intact subjects and two transfemoral amputees participated in this study for the quantification of CES parameters. One of the metrics used to evaluate the performance of CES was the symmetry index.

- Symmetry index (SI)

  For measuring the gate performance of lower limb amputees, the symmetry was quantified during stance and swing by symmetry index based on the following equation:

$$SI = \frac{(S - P)}{(S + P) \times 0.5} \qquad (22)$$

where $S$ is the gait phase duration of the subject's sound and $P$ is the gait phase duration of prosthetic lower limbs.

In 2017, Xia *et al*. [163] proposed a combinational method of recurrent neural networks (RNNs) and convolutional neural networks (CNNs) to estimate kinematics from multi-channel EMG signals. Five-channel EMG electrodes and the motion tracker system were used to record data from eight able-bodied subjects. The upper limb movements were analyzed in 3D space, and the R-squared metric evaluated the method's performance.

- R-squared ($R^2$)

  $R^2$ is the ratio between the sum of squared error from the estimated 3D path and the actual path's total square residual.

$$R^2 = 1 - \frac{\sum_{t=0}^{T}(\hat{x}_t - x_t)^2 + (\hat{y}_t - y_t)^2 + (\hat{z}_t - z_t)^2}{\sum_{t=0}^{T}(x_t - \bar{x}_t)^2 + (y_t - \bar{y}_t)^2 + (z_t - \bar{z}_t)^2} \quad (23)$$

  where $T$ is the length of the time series and $[x_t, y_t, z_t]$, $[\hat{x}_t, \hat{y}_t, \hat{z}_t]$ and $[\bar{x}_t, \bar{y}_t, \bar{z}_t]$ are the actual position, estimated position, and the mean position vector, respectively.

In 1999, Lai *et al*. [164] experimented with evaluating an artificial neural network (ANN) to estimate joint torque in isokinetic movements. sEMG signal was recorded by eight pairs of electrodes from the arm of five healthy subjects reclined on the bed with a fixed arm to perform elbow joint motions. The root-mean-square difference and cross-correlation were defined to analyze the ANN model's performance and the effect of cross-talk, respectively.

- Root-mean square difference (RMSD)

  The root-mean-square difference of the measured torque $\tau_a(i)$ and the estimated torque $\tau(\mathrm{i})$ was computed as below:

$$RMSD = \sqrt{\frac{\sum_i \left(\tau(\mathrm{i}) - \tau_a(i)\right)^2}{\sum_i \left(\tau_a(i)\right)^2}} \quad (24)$$

- Cross-correlation

  The effect of using different electrode sites for recording EMG signal is evaluated by the cross-correlation between any two EMG signals such as $x(n)$ and $y(n)$ by the following equation:

$$R_{xy}(n) = \frac{1}{N}\sum_{i=0}^{1} x(\mathrm{i})y(\mathrm{i} - n) \quad (25)$$

  where $N$ is the length of records.

## 4. Discussion

Various schemes for controlling prosthesis hands were reviewed. A drawback of the proportional and direct myoelectric control is that the number of movements is confined to the number of sEMG channels, not a problem for advanced controls such as pattern recognition and regression methods [57]. Although pattern recognition and deep learning approaches bring their benefit to the field, there are still many challenges to resolve. Some prosthetic upper limbs can interact with the elbow, and patients may suffer from discomfort, sores, and blisters, while others experience heat, sweating, and chafing [165]. The identification of opening and closing positions of the hand from the residual limb is still tricky. Residual biceps and triceps are mainly used for hand prostheses. However, they do not fully show the details for closing and opening the hand [166]. On the hardware side, most of the upper limb prostheses are heavy, have short battery life, and do not have a realistic appearance. In addition to the training requirements per user and their high cost, some more technical challenges make it less attractive.

The sEMG signal is often recorded while the user is in a static position (sitting), but prosthesis users have to use the system in multiple positions in a real-time scenario (e.g., walking, climbing stairs), which influences the efficiency of the system [167]. External interferences, a shift in electrode impedance, muscle weakness, and electrode movement are other frequent issues that deteriorate functional usage [168]. Electrode displacement happens every time electrodes are marginally reconciled at a different location than the underlying musculature as users use a prosthesis. When the user completes an activity, a displacement of the electrodes happens. Such an electrode shift will lead to a difference in the limb's EMG characteristic (recording), making it harder to decipher the gestures [169]. The same limb supports different muscle contraction forces in diverse environments when conducting regular activities. Therefore, owing to the same targeted limb, the difference in muscle contraction intensity results in myoelectric signal pattern classification uncertainty that impacts the prosthesis's performance [170].

## 5. Conclusions

In this paper, we provided a comprehensive review of myoelectric prosthesis control. Some global statistics were first provided for the unilateral and bilateral upper-limb amputations. The technical terms' terminology was then provided, and the following myoelectric control methods were discussed in detail: On-off and finite-state, proportional, direct, and posture, simultaneous, classification and regression-based control, and deep learning methods. Myoelectric control Performance indices were reviewed, and finally, the advantages and disadvantages of the control methods were discussed.

**Author Contributions:** MRM, MN, FZ, ZNE, GA, NM, FY, MR, and HRM participated in the Conceptualization, Investigation, and Methodology. MRM and HRM participated in the Project administration. MRM, MN, FZ, ZNE, GA, NM, FY, and MR participated in the literature review. MRM and HRM participated in the interpretation of the results and supervision. MRM, MN, FZ, ZNE, GA, NM, FY, and MR participated in Writing – original draft, while MRM and HRM participated in Writing – review & editing. All authors read and approved the final manuscript and agreed to be accountable for all aspects of the work.

**Funding:** This research received no external funding.

**Institutional Review Board Statement:** Not applicable.

**Informed Consent Statement:** Not applicable.

**Data Availability Statement:** Data supporting the reported results can be found at PubMed (https://pubmed.ncbi.nlm.nih.gov) and Cochrane Library (https://www.cochranelibrary.com).

**Conflicts of Interest:** The authors declare no conflict of interest.